\documentclass[osajnl,10pt,letterpaper]{article}
\usepackage{opex3}

\begin{document}

\title{Single and dual fiber nano-tip optical tweezers: trapping and analysis}

\author{Jean-Baptiste Decombe,* Serge Huant, and Jochen Fick}

\address{Univ. Grenoble Alpes, Inst NEEL, F-38042 Grenoble, France\\
        CNRS, Inst NEEL, F-38042 Grenoble, France}

\email{*jean-baptiste.decombe@grenoble.cnrs.fr} 



\begin{abstract}An original optical tweezers using one or two chemically etched fiber nano-tips is developed. We demonstrate optical trapping of 1 micrometer polystyrene spheres at optical powers down to 2 mW. Harmonic trap potentials were found in the case of dual fiber tweezers by analyzing the trapped particle position fluctuations. The trap stiffness was deduced using three different models. Consistent values of up to 1 fN/nm were found. The stiffness linearly decreases with decreasing light intensity and increasing fiber tip-to-tip distance. 
\end{abstract}




\section{Introduction}
 Optical tweezers are now well-established since the pioneering work of A. Ashkin \cite{ADB+86}. This non-contact technique is of large interest in many scientific domains such as biochemistry, physics and medicine. The application of optical near-field tweezers allows to lower the trapping intensities or to trap smaller particles. Cavities of dielectric \cite{HLS+12} or photonic waveguides \cite{RDC+12} have been used to collect and trap micro-particles. Exploiting the strong optical field gradient of surface plasmon cavities allowed the realization of stable nanoparticle trapping. Different geometries such as  strips \cite{WSS+10}, dipole antennas \cite{ZLS+10}, nano-block pairs \cite{TS11} or double nano-holes \cite{PG11} were successfully applied.

The use of optical fibers attracts increasing attention as highly flexible tools for particle trapping. Fiber-based optical tweezers do not require substrates or bulky high numerical aperture objectives. They provide easy access to the trapped particle, which is useful for the implementation of further manipulation or characterization elements. 
Examples of tweezers with two facing optical fibers includes micro-fluidic actuators using two cleaved fibers \cite{BLM12}, tweezers based on fiber tips grown by photo-polymerization \cite{VOO09} or using tapered lensed fiber tips \cite{LS95}. 

Single fiber tip tweezers were realized with a micro-lensed cleaved fibers \cite{NTO+06}, a multicore lensed fiber \cite{BKA+13}, single- or multi-mode chemically etched fiber tip \cite{LGY+06,LWL+13} or by gradient index chemically etched fiber tips creating 3D bottle beams \cite{MPK12}.

The measurement of the optical forces acting on the trapped particles is of paramount interest. It allows to determine the trapping efficiency. Moreover, due to the very weak optical forces, it can be used for high sensitivity force or displacement sensing \cite{MHN+10}. Different theoretical approaches allow to determine the optical forces of an optical tweezers from the trapped particle position fluctuations. In \cite{BF04}, K. Berg-S{\o}rensen and H. Flyvberg present an overview of these models. High frequency quadrant positions sensors are currently used for particle position recording. It was, however, shown that CMOS camera videos with frame rates of some hundred Hz are sufficient for the accurate force measurements \cite{GLK+08}. 

In the present paper, we report on optical trapping of micrometer size dielectric particles using one or two bare optical fiber nano-tips. These chemically etched fiber tips with nanometer size apex were already used for high resolution optical scanning microscopy \cite{DSV+11}. The trapping efficiency at different light powers and fiber distances is evaluated by analyzing the experimental data within three different models that find very consistent results. 
 
\section{Experimental}
\begin{figure}
	\centering
	\includegraphics[width=5.cm]{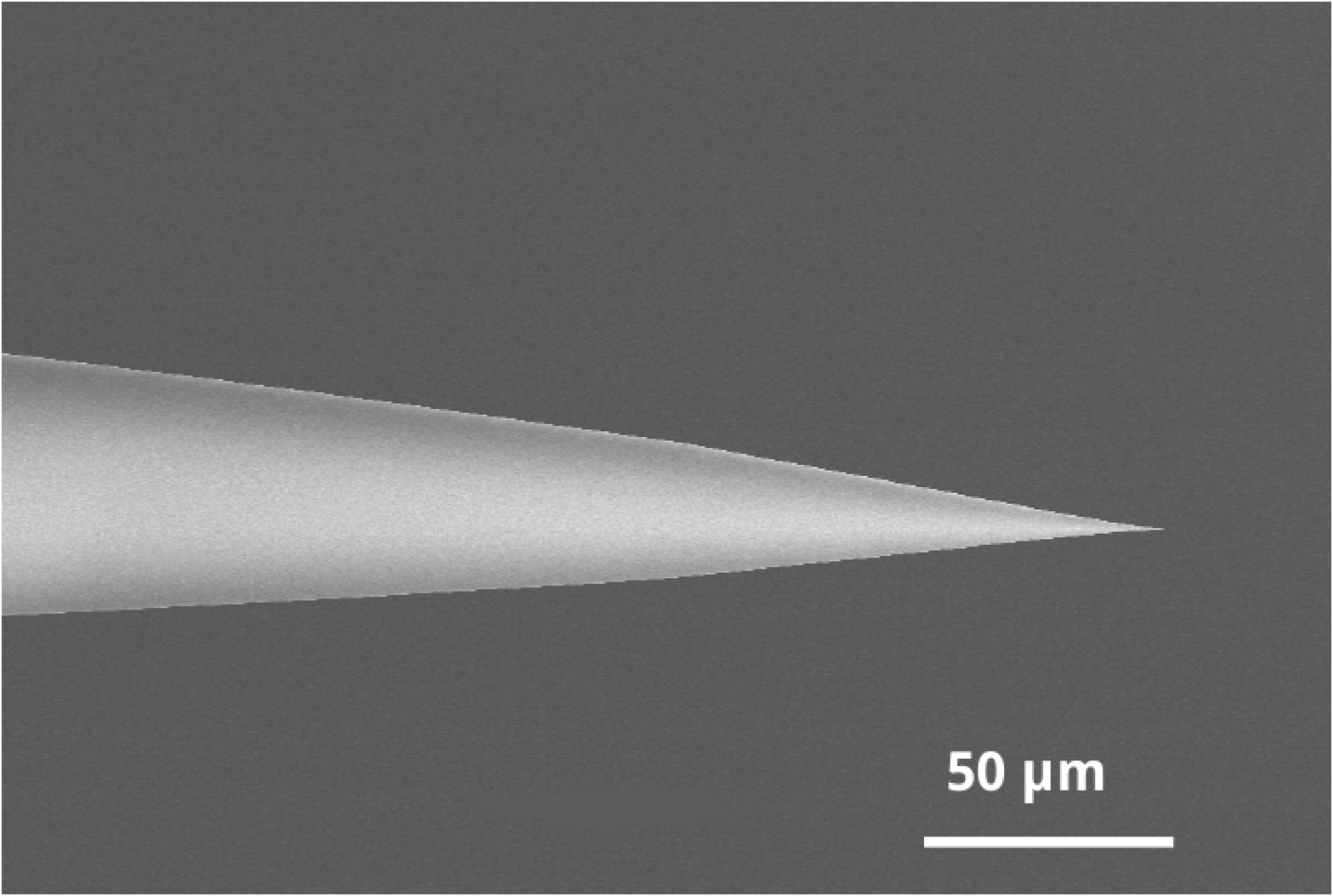}\hfil
	\includegraphics[width=5.cm]{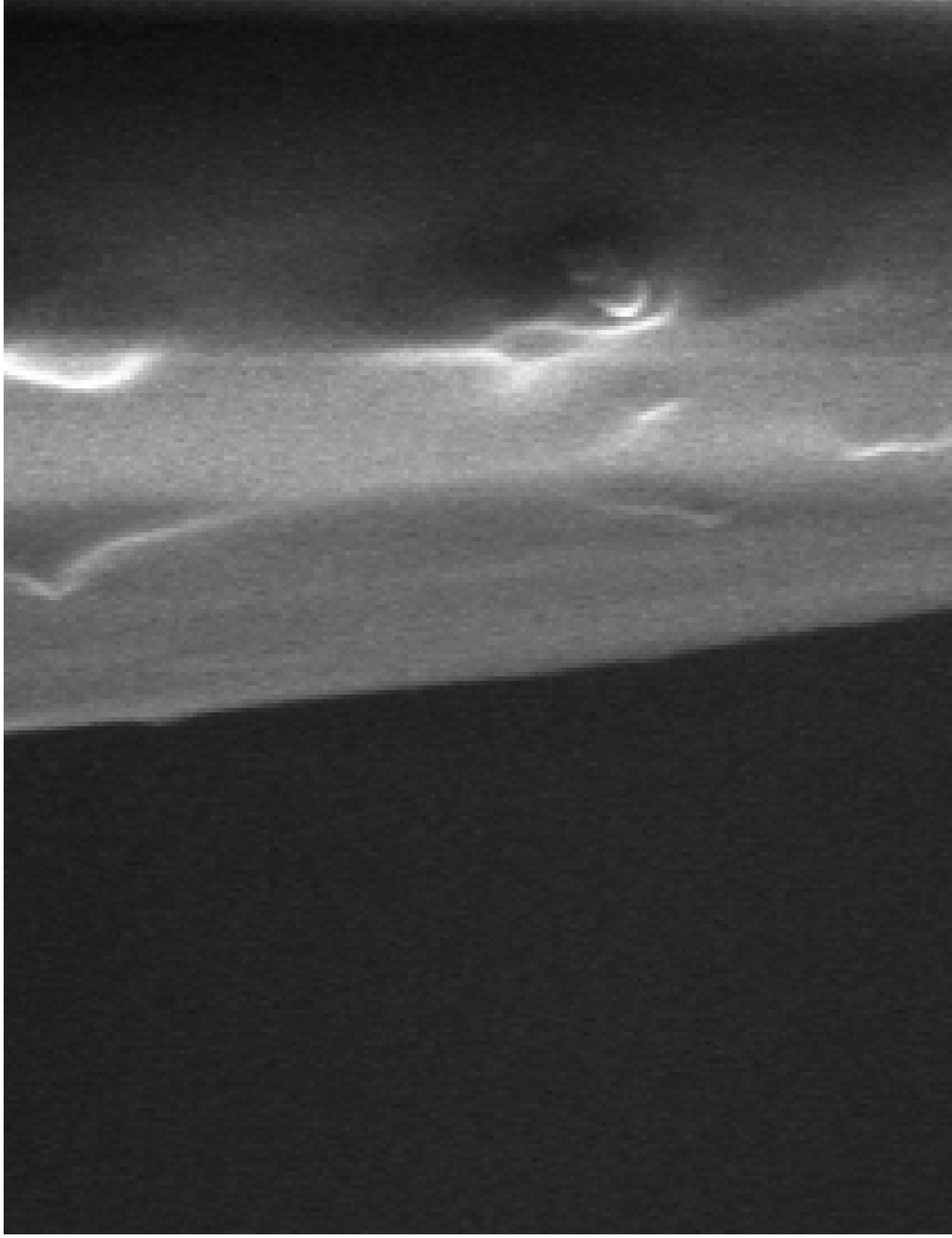}
	\caption{Scanning electron microscope images of an etched fiber tip.\label{fig.SEM}}
\end{figure}

Optical fiber tips are elaborated by chemical etching in aqueous hydrofluoric acid of standard pure silica core single mode optical fibers (S630-HP, Nufern) ~\cite{CSM+06}. The obtained fiber tips are reproducible with smooth surfaces, full angle of about $15^{\circ}$, and apex diameters of 60~nm  [Fig. \ref{fig.SEM}]. 

\begin{figure}[b]
\centering
	\includegraphics[width=10.cm]{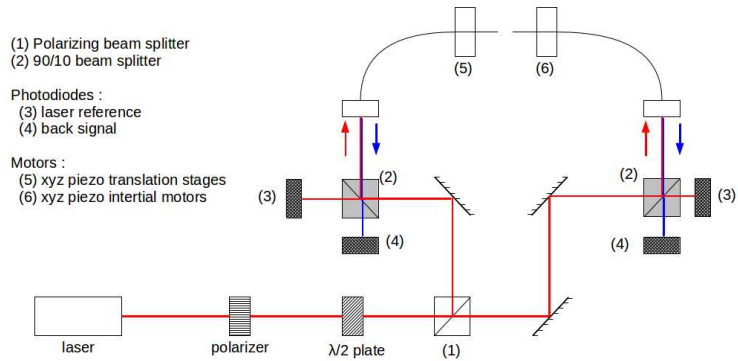}\\
	\caption{Scheme of the experimental set-up. \label{fig.setup}}
\end{figure}

The scheme of the optical tweezers set-up is shown on Fig. \ref{fig.setup}. A 808~nm single mode pigtailed laser diode with a maximum output power of 250 mW (LU0808M250, Lumics) is linearly polarized and split by a polarized beam splitter. The relative intensity in the 2 arms is controlled by an half-wave plate. In each arm, the beam is split again with a 90/10 beam splitter. 90\% of the light is coupled into the optical fiber tips whereas the other 10\% go directly in a photodiode to give the laser reference signal. The reflected intensity is measured by an amplified Si-photodiode (New Focus 2001) placed after the beam splitter. This "back signal" is a superposition of the transmitted light from the second, opposing fiber tip and the reflection from the tip and, if applicable, the trapped particle. 

Fiber tips are mounted on two sets of $xyz$ translation stages : piezoelectric translation stages (PI P620) with sub-nanometer resolution and $50~\mu$m range, and inertial piezoelectric translation stages (Mechonics MS 30) with $\approx 30$ nm step size and up to 2.5~cm range. Transverse transmission intensity maps are recorded by scanning one fiber tip in a plane perpendicular to the tip axes. The entire set-up is controlled by homemade software in the LabView environment.

A homemade microscope with a $\times50$ objective (Mitutoyo) coupled to a CMOS camera (Thorlabs) is used for visualizing the trapped particles. The objective long working distance (13.9 mm) allows easy particle observation. The main drawbacks are, however, its low resolution of about 0.5 $\mu$m and its small focus depth of 1.1 $\mu$m. The camera allows a full frame size of $1280 \times 1024$ pixels with 58 nm/pixel resolution and frame rates of 11 fps. The frame rate can be boosted to 115 fps by reducing the frame  size to $130 \times 90$ pixels. For higher frame rates the limited CMOS sensitivity results in low contrast images. 

The fluid chamber consists of an o-ring placed in between two glass slides and cut in two parts in order to insert the fiber tips. The system is sealed using vacuum grease allowing to work several hours in a stable system, without evaporation. The chamber is fixed on a set of translation and rotation stages to allow easy alignment in respect to the fiber tips. 
Commercial 1 $\mu$m polystyrene spheres (Corpuscular) are used for trapping in aqueous suspension with 0.125 mg/mL concentration.

Light power at the end of the fiber tips is measured by means of a power meter. Transmission measurements between two identical fiber tips indicate that the fiber-to-fiber transmission is about 3 times more efficient in water than in air. Thus, for the trapping measurements a correction factor of 1.73 is applied to the measured intensity. All values given in this paper are measured and corrected intensities at the end of one single fiber tip.

An automatic particle tracking software is developed in the open source  Scilab environment in order to exploit trapping videos with more than $3\times 10^5$ frames. The program detects the center of the particle surface in each frame. The estimated resolution of 50 nm is of the order of the pixel size, but below the microscope resolution. 

\begin{figure}
	\centering
	\includegraphics[width=6.cm]{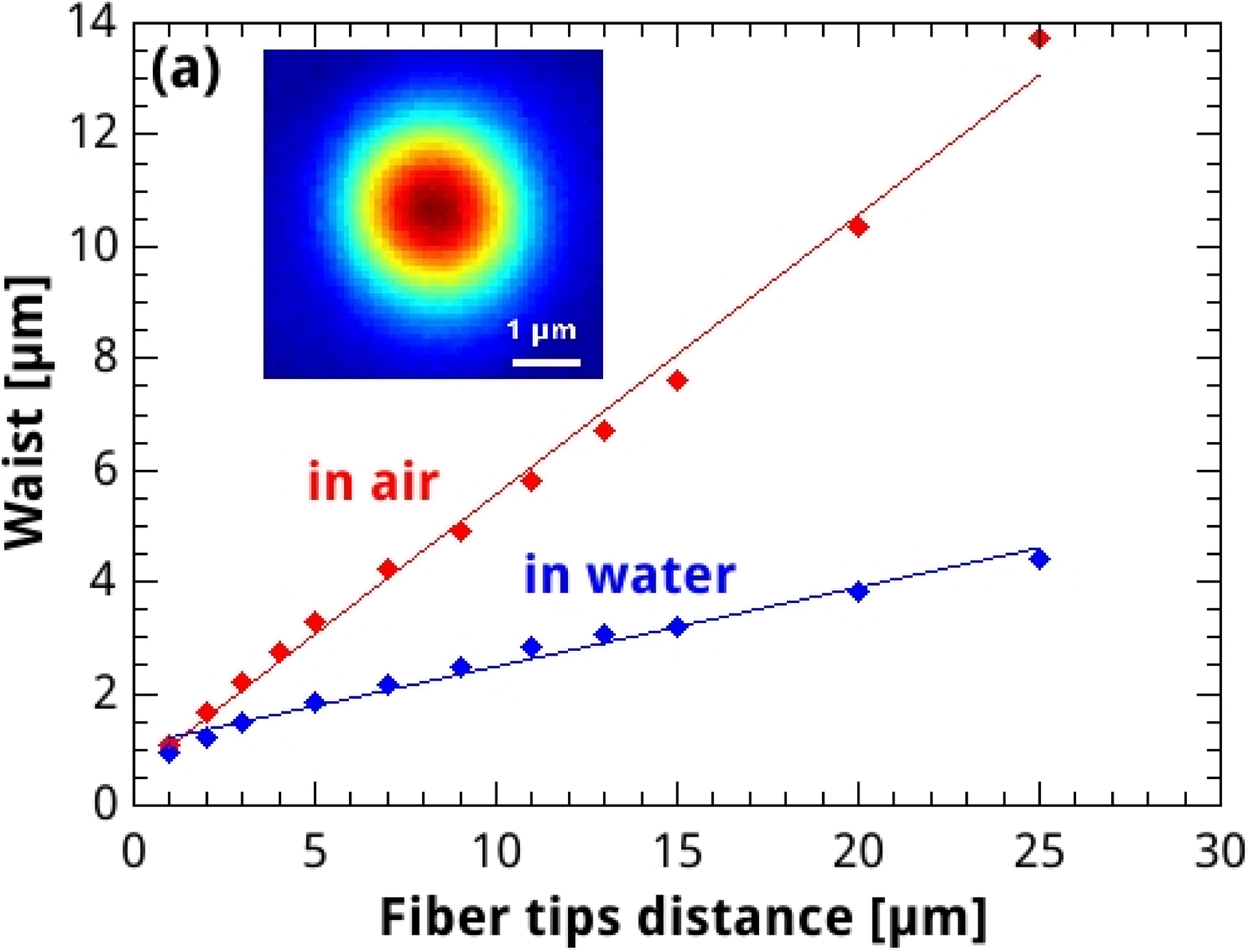}
	\includegraphics[width=6.cm]{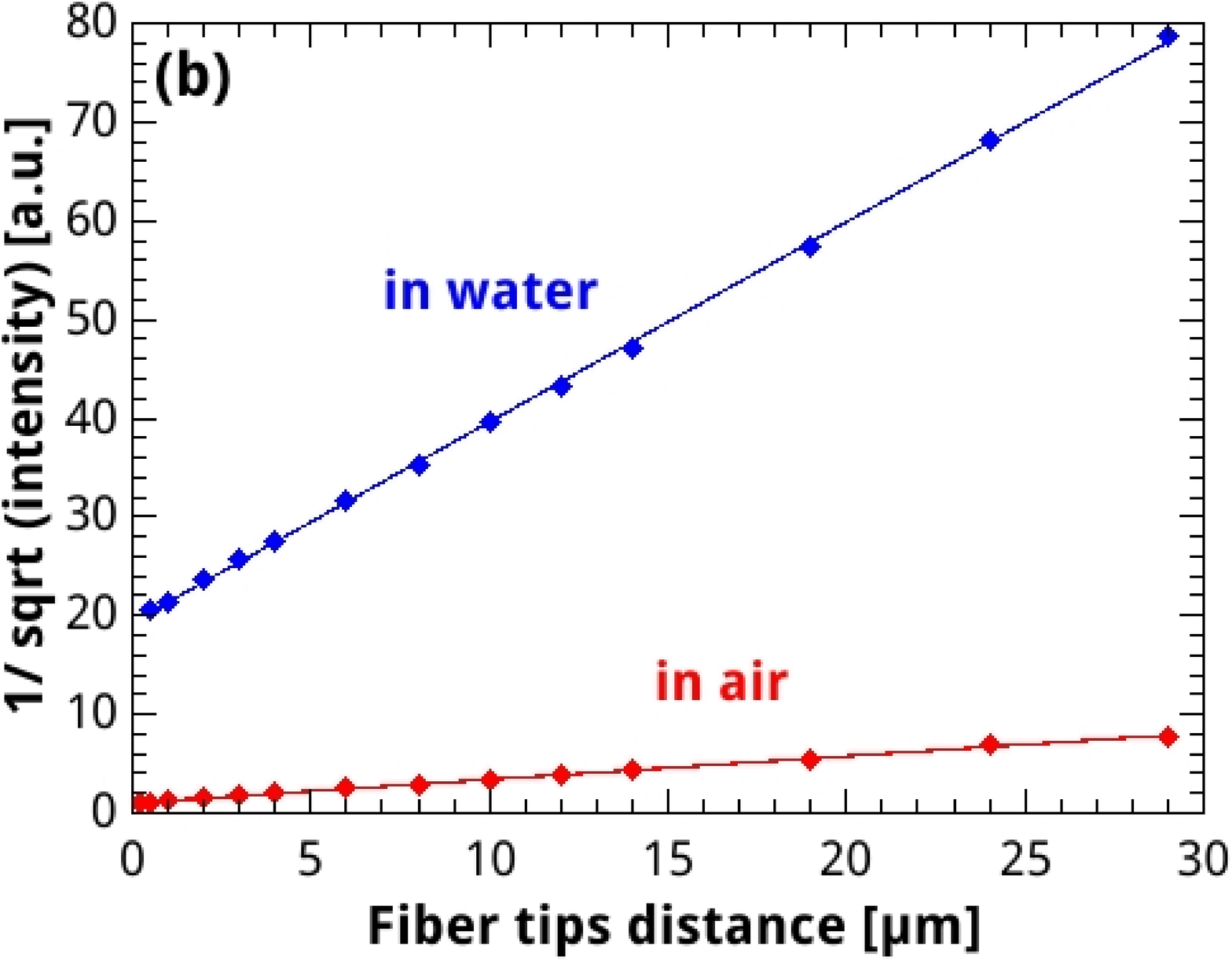}
	\caption{Transmission spot width/ waist (a) and intensity (b) as a function of fiber tip-to-tip  distance. Insert : transverse transmission intensity map in air ($d=1~\mu$m). \label{fig.waist}}
\end{figure}

\section{Results and discussion}
\subsection{Optical fiber tip emission properties}

The light emission properties of the fiber tips are studied by transverse transmission maps. In the measurement, the laser beam is injected in one fiber and the transmitted light is collected by scanning the second, identical fiber tip in the transverse plane. The obtained transmission maps show circular  Gaussian-shaped intensity spots [insert Fig. \ref{fig.waist}(a)]. 

To obtain the emission spot width of a single fiber tip, the recorded transmission maps have to be corrected by the capture function of the second fiber tip. For two identical fiber tips with Gaussian emission profiles the corrected waist $w$ is calculated by  $w=(\tilde w^2-\frac{1}{2}w_0^2)^{1/2}$ with $\tilde w$ the as-measured waist and $w_{0}$ the measured waist at smallest tip-to-tip distances \cite{DBM+13}.

Measurements are performed for tip-to-tip distances between 1 and 25~$\mu$m in air and in water [Fig. \ref{fig.waist}]. At smallest distances, the minimal measured waists are $\sim$900~nm in both media. As shown previously, sub-wavelength spot sizes can only be obtained with metalized fiber tips \cite{DBM+13}. The actual beam size is, however, of the same order that the trapped particles. The waist increases linearly with tip-to-tip distance. In air, the full angle beam divergence is 30$^{\circ}$, corresponding to a numerical aperture of 0.25. This result is confirmed by far-field angular measurements  using an optical goniometer. Due to the higher refractive index  the beam is much less divergent in water. In this case, the emission angle is $\sim$ 8$^{\circ}$ with N.A. = 0.07. 

The transmission intensity is decreasing with fiber tip-to-tip distance with a linear behavior in the $I^{-1/2}$ plot. As already stated, the intensity is about three times higher in water than in air.

\begin{figure}
\centering
	\includegraphics[width=7.cm]{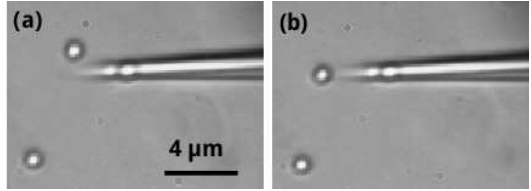}\\
	\caption{Optical microscope images of a trapped 1~$\mu$m polystyrene sphere. (a) Laser off. (b) Laser on (9 mW) (\url{Media 1}). \label{fig.t1f}}
\end{figure}

\subsection{Single fiber tip tweezers}

Transient trapping for few minutes of 1~$\mu$m spheres is obtained with one single fiber tip. In this configuration the attractive optical gradient force ($F_{grad}$) attracts the particle towards the optical axes at the end of the fiber tip [Fig. \ref{fig.t1f} and \url{Media 1}]. Because of the Brownian motion the particle can temporarily leave this stable trapping position and be ejected by the repulsive optical scattering force ($F_{scat}$).

\begin{figure}[b]
\centering
	\includegraphics[width=6.cm]{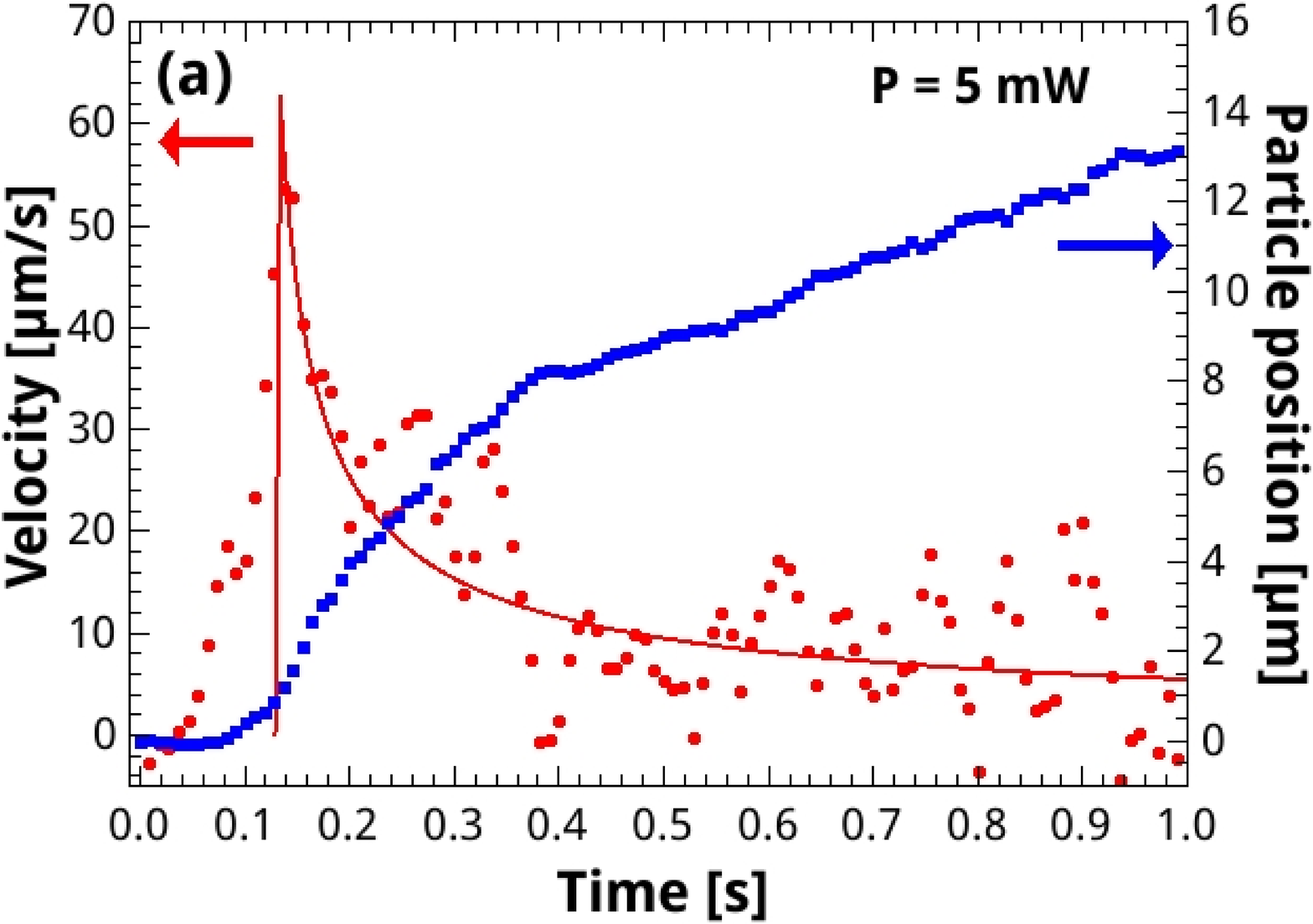}
	\includegraphics[width=6.cm]{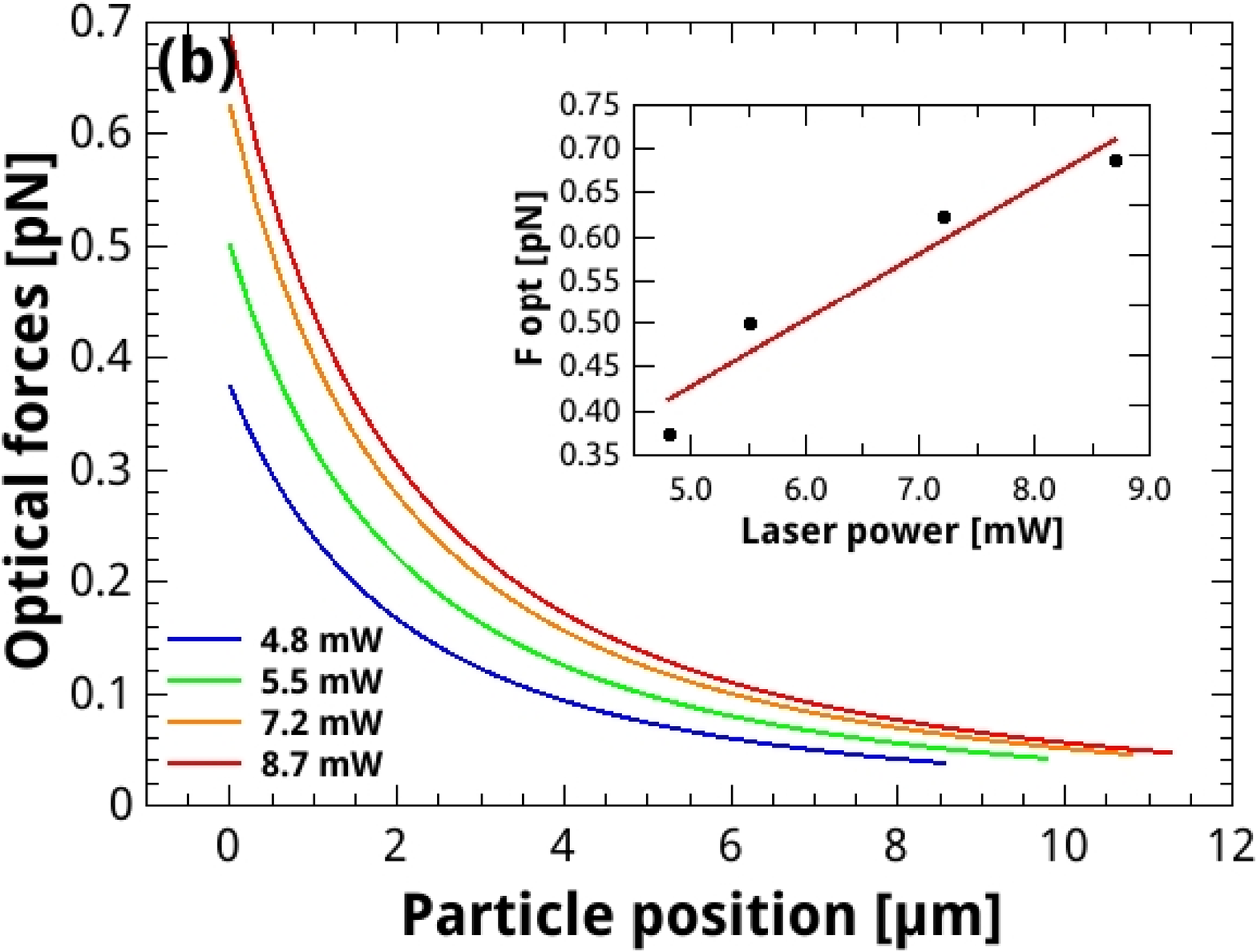}\\
	\caption{(a) Evolution of particle position (blue dots) and velocity (red). The line corresponds to the calculated velocity curve. (b) Optical force as a function of particle position for 4 different laser powers. The inset shows the optical force as a function of laser intensity at a particle position of 0$~\mu$m . \label{fig.t1ff}}
\end{figure}

The observation of the particle speed during ejection allows to deduce the optical forces using Newton's second law : 

\begin{equation}\label{eq.newton}
 F_{opt}+F_{stokes}=m\dot{v}
\end{equation}

with $m$ and $v$ the particle mass and speed. The Stokes force is defined by $F_{stokes}= -\gamma_0 v$, with $\gamma_0=6\pi a\eta$ the friction coefficient, $a$ the sphere radius, and $\eta$ the dynamic viscosity of water. $F_{opt}= F_{grad}+F_{scat}$ is the total optical force. In the case of particle ejection we suppose that $F_{grad}\ll F_{scat}$. The scattering force is scaling with the light intensity which itself is decreasing by ${1/x^2}$ with tip-to-tip distance [Fig. \ref{fig.waist}(b)]. Now, Eq. \ref{eq.newton} becomes  $m\ddot x +\gamma_0 \dot v - F_0/x^2=0$ with $F_0$ the maximal optical force, which is the only free parameter. This equation can be solved by numerical differential equation solvers.

Particle ejection events are observed for different light intensities. The particle position and speed as a function of time are obtained from the recorded videos using our tracking software [Fig. \ref{fig.t1ff}(a)]. The particles are first accelerated by the strong optical forces, before being slowed down by the Stokes drag. Finally, particles comes to a standstill as the optical force becomes negligible.

$F_0$ is determinate by adjusting the calculated velocity curve to the experimental data. The experimentally observed  particle acceleration is slower than predicted by our model. This is mainly due to the neglect of the gradient force near the fiber tip. The accordance with theory is good during the particle deceleration phase. The deduced optical forces are in the 0.4 -- 0.7 pN range. They are linearly scaling with the light intensity [Fig. \ref{fig.t1ff}(b)].

Using one fiber tip, spheres can only be trapped very close to the tip apex where the gradient force dominates the repulsive scattering force. Thus the particle can stick to the fiber due to strong Van der Waals forces. In the present case particles leave the fiber tip shortly after the laser is switched off.

\subsection{Dual fiber nano-tips tweezers}

1~$\mu$m spheres are also stably trapped between two fiber tips. In the dual-beam configuration when the fibers are well aligned coaxially, the particle is confined on the common optical axis by the transverse gradient forces and stabilized at a point on the optical axis where the two repulsive scattering forces are canceled.

\begin{figure}
\centering
	\includegraphics[width=4.5cm]{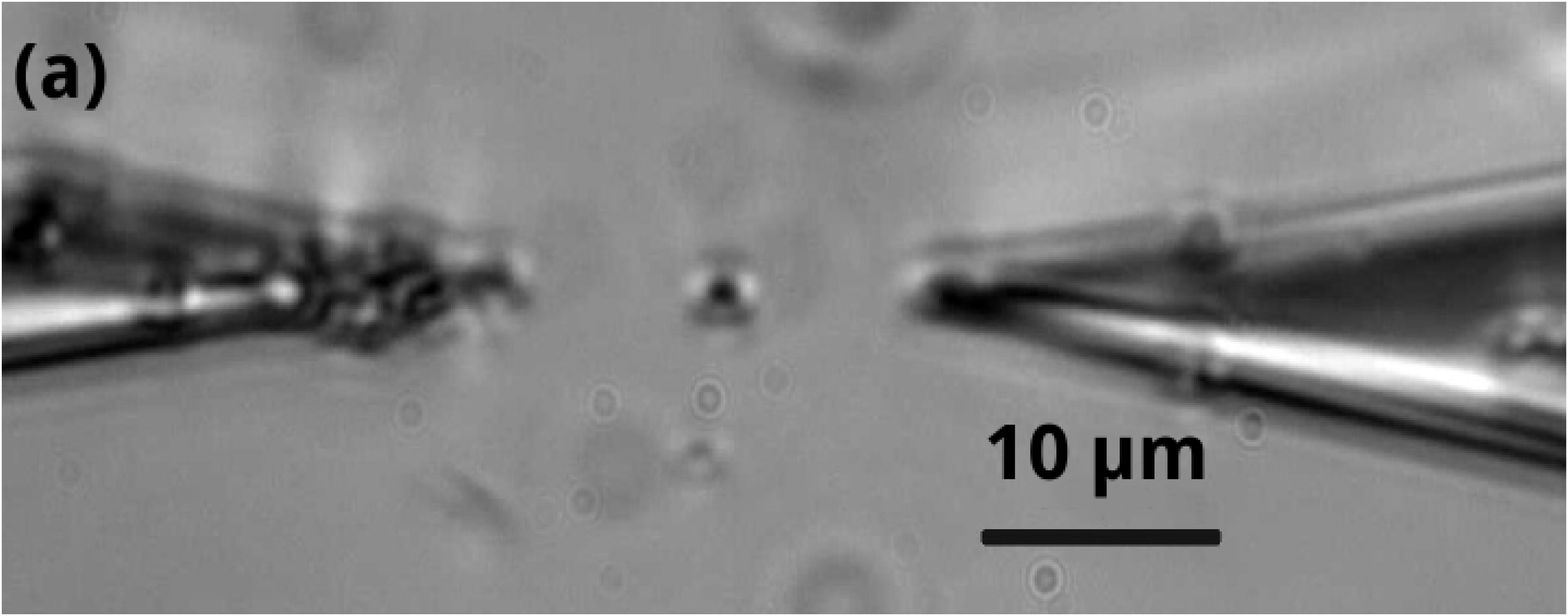}
	\includegraphics[width=4.5cm]{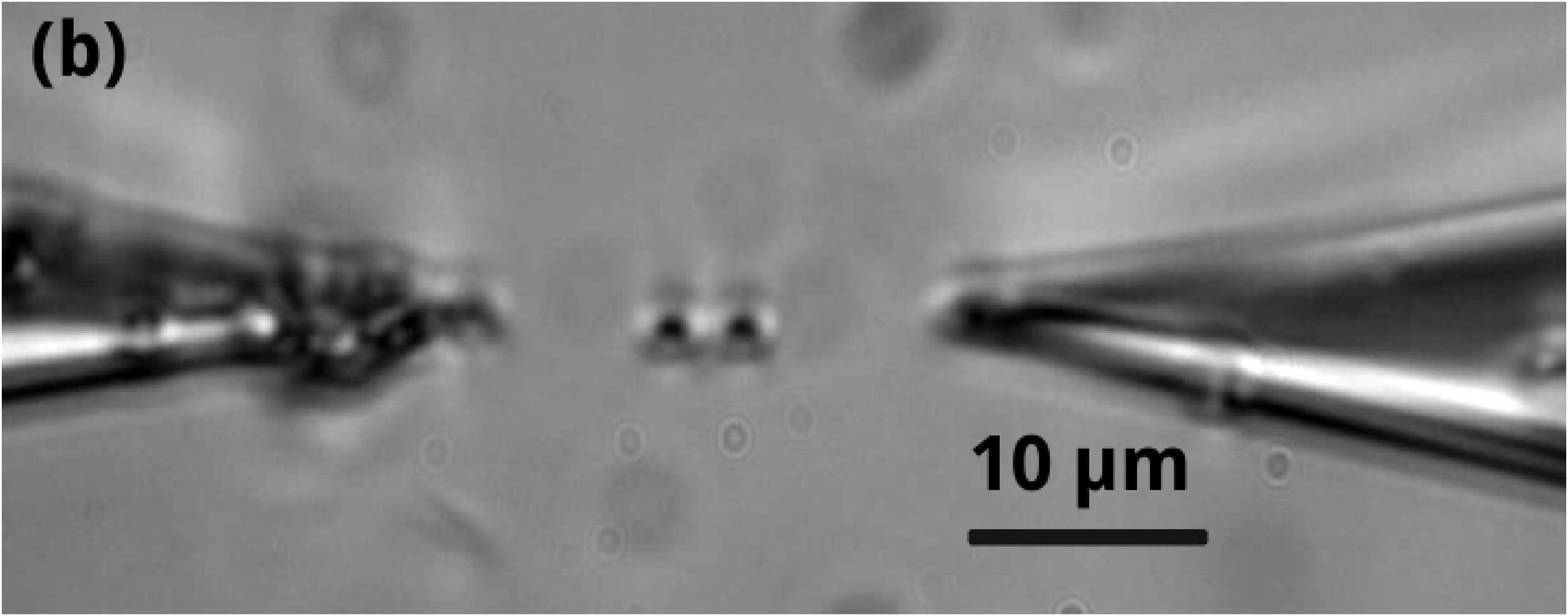}\\
   \includegraphics[width=4.5cm]{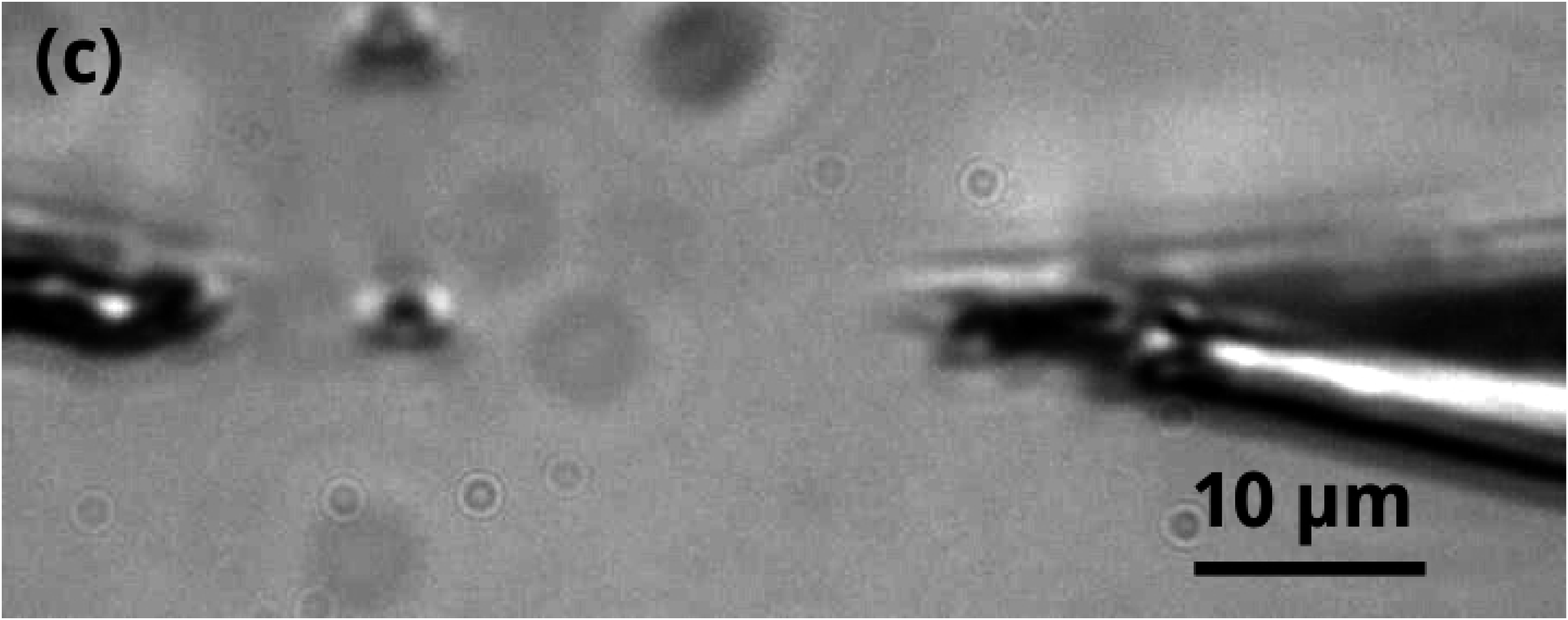} 
	\includegraphics[width=4.5cm]{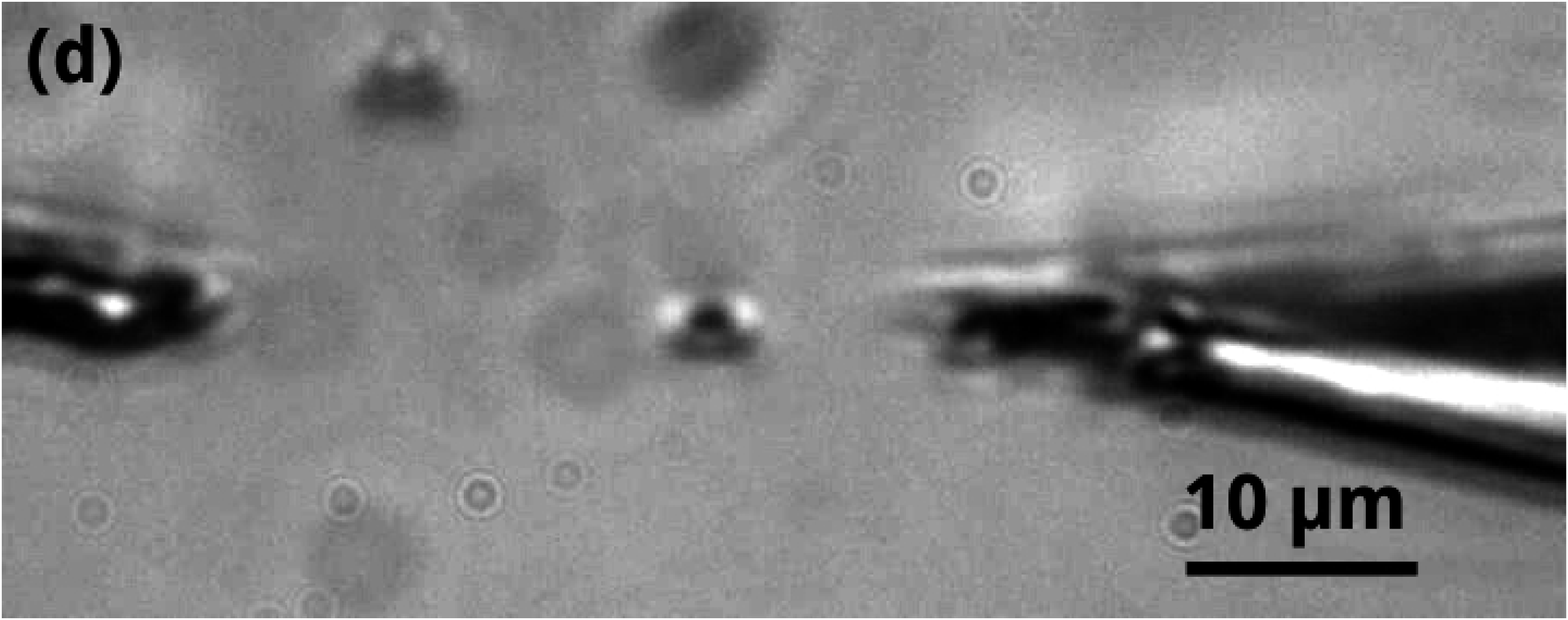}\\

	\caption{Trapping of one (a) and two (b) spheres with 2 fiber nano-tips (\url{Media 2}) ($I=6$ mW, $d= 17~\mu$m). (c)-(d) Control of the particle position by modifying the relative light intensities injected in the two facing nano-tips.  \label{trap}}
\end{figure}

Experimentally, fiber tips are aligned by scanning one fiber relative to the second one as previously discussed.  Then they are placed at the maximum transmission intensity position. Stable trapping for several hours is observed for one sphere with light power of 3.5 - 10 mW and fiber tip-to-tip distances of 6 to 17~$\mu$m [Fig. \ref{trap}(a)].

The sphere position along the optical axis can be controlled by varying the relative optical intensity injected into both fibers. By increasing the intensity in the left hand side fiber, the  trapped particle moves to the right [Figs. \ref{trap}(c) and \ref{trap}(d)].
We were also able to trap 2 and 3 spheres at the same time during few minutes [see Fig. \ref{trap}(b) and \url{Media 2}].

\begin{figure}
	\centering
	\includegraphics[width=6.cm]{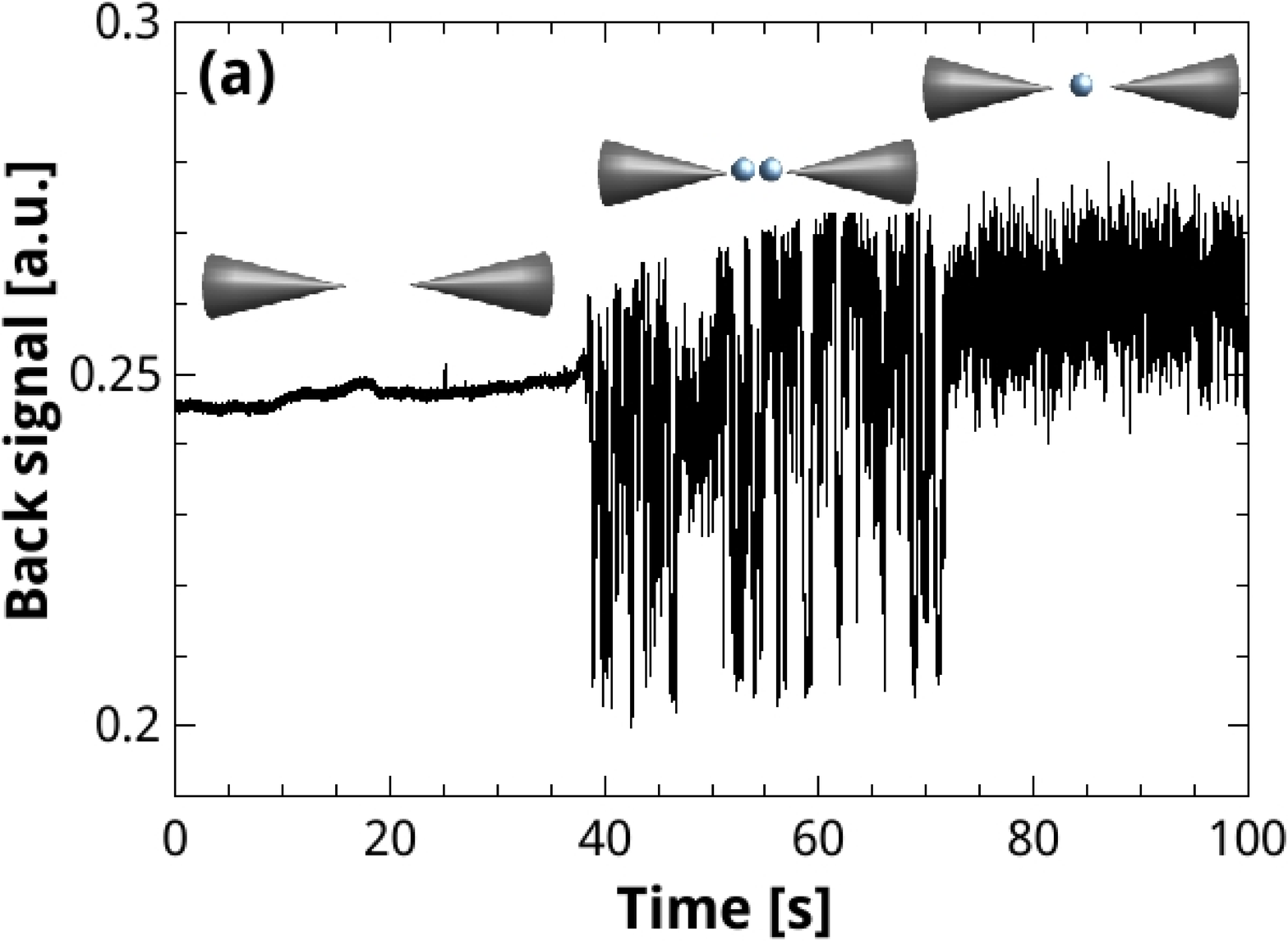}
	\includegraphics[width=6.cm]{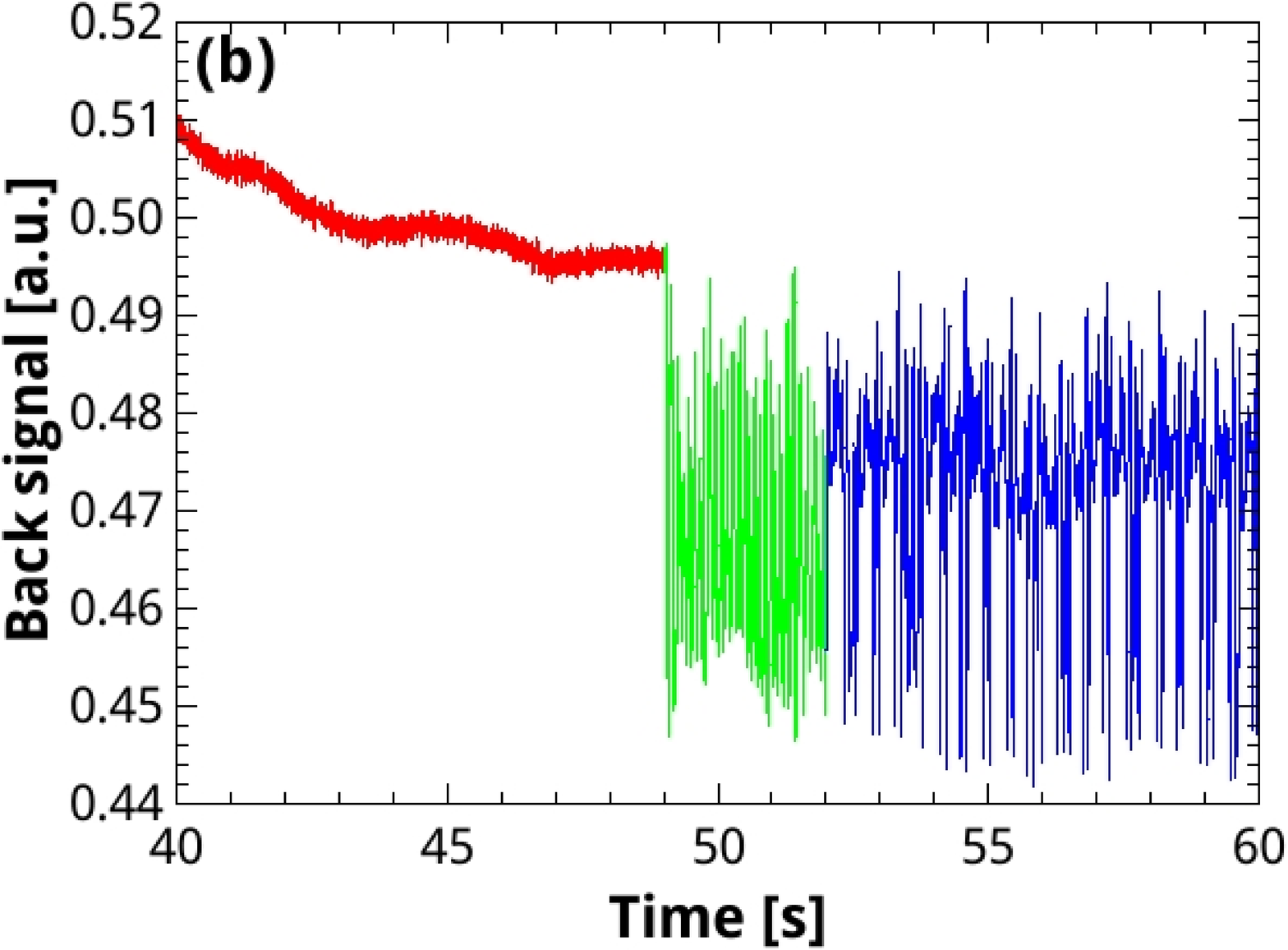}\\
	\includegraphics[width=6.cm]{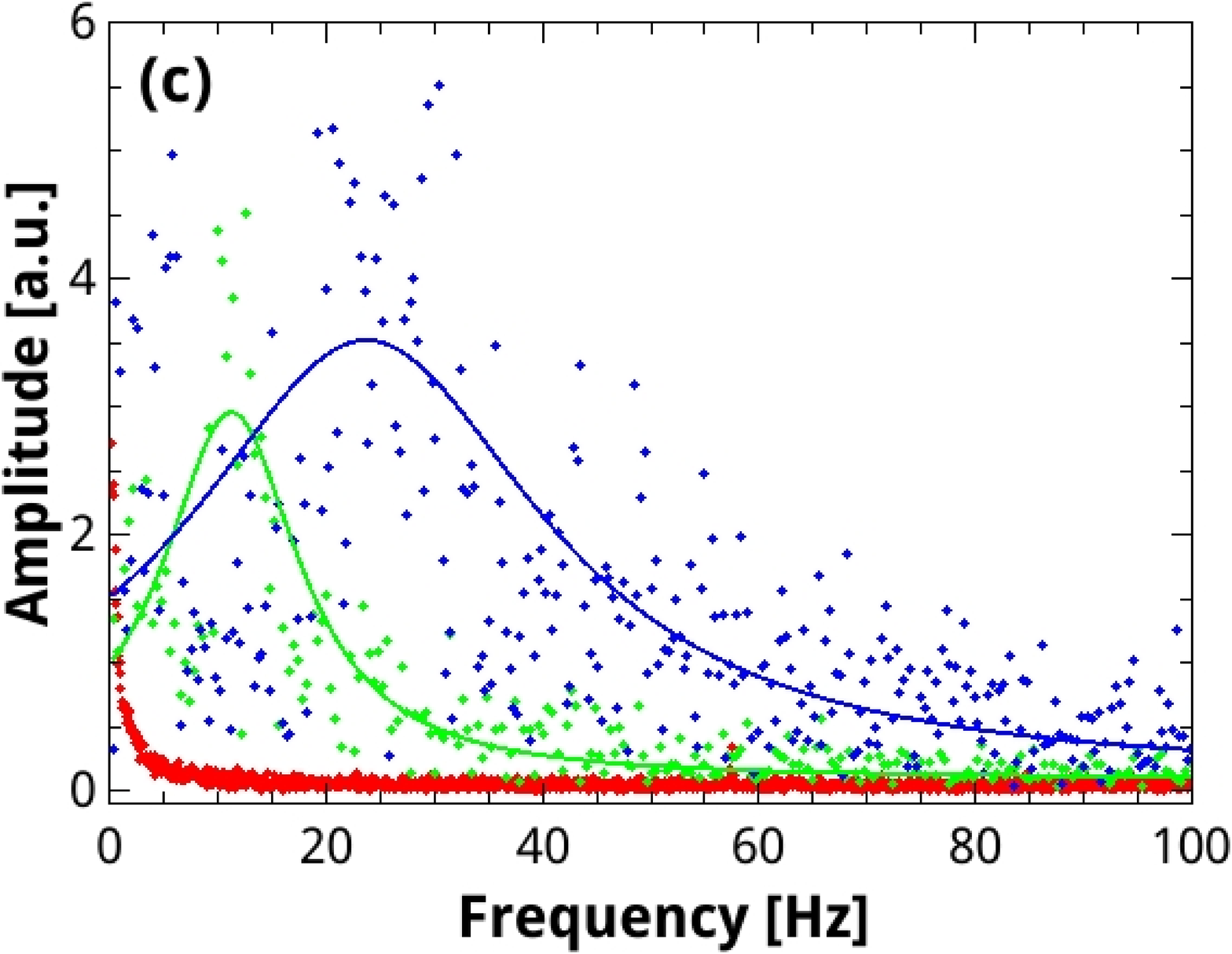}
	\includegraphics[width=6.cm]{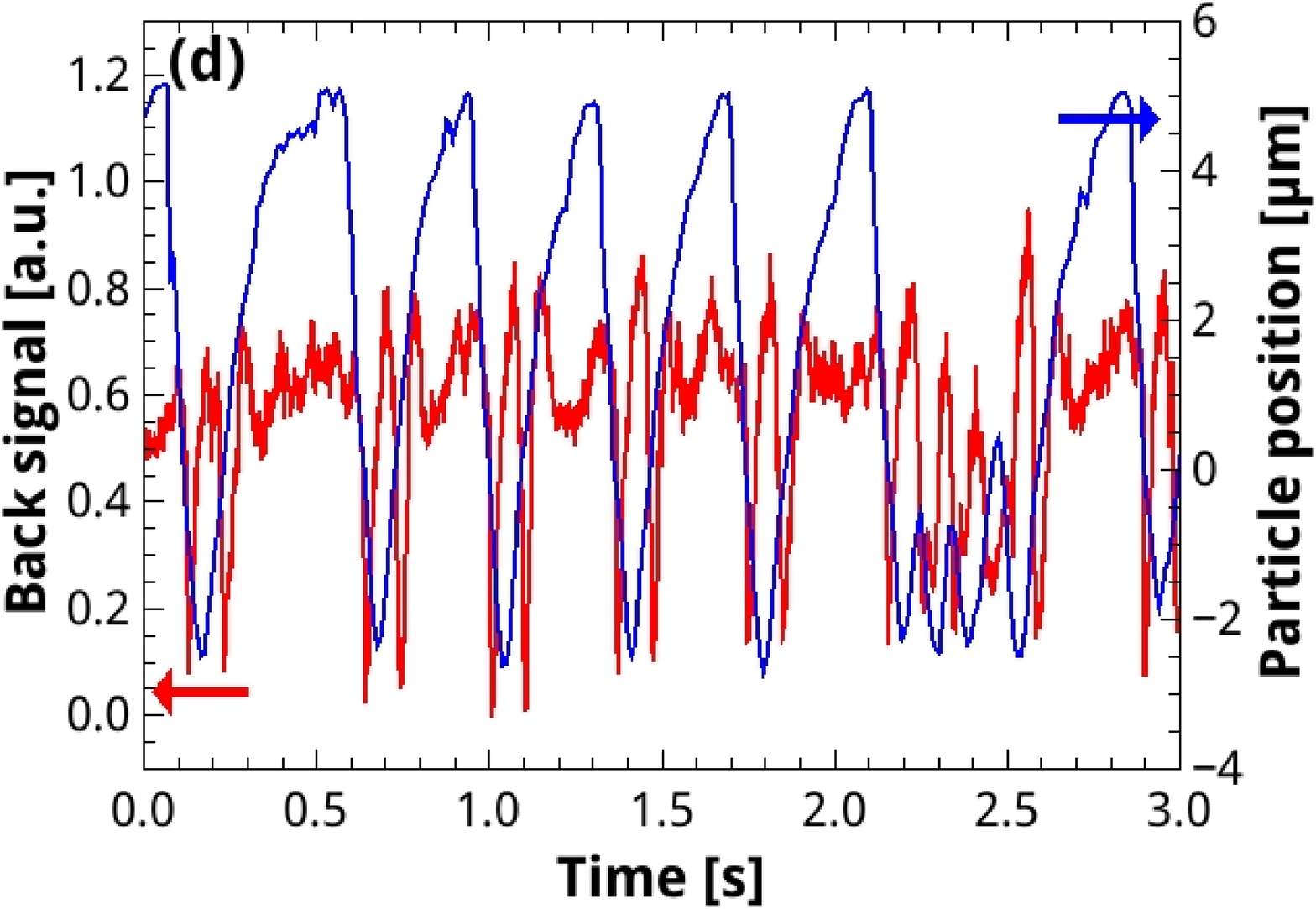}
	\caption{Back signal as a function of the trap state: (a) zero, two, and one trapped particle. (b) No particle (red), stable trapping (green), and metastable trapping (blue). (b) Fast Fourier Transform of these 3 states (lines are Lorentzian best fits). (d) Comparison between back signal (red) and particle position (blue) for metastable particle trapping (\url{Media 3}).
	\label{fig.back}}
\end{figure}

Besides the observation by the microscope, the back signal reflected by the  fibers tips gives valuable information about the trapping events [Fig. \ref{fig.back}]. Particle trapping results in significant intensity modulations. This effect is even more pronounced for trapping of multiple particles. 

The relation between the back signal and particle trapping can be investigated with a particle which is trapped for some seconds in the central position between the fibers before oscillating between two metastable positions [Figs. \ref{fig.back}(b)--\ref{fig.back}(d)] and \url{Media 3}). In the case of stable trapping strong oscillations with main frequency peaks at 10 Hz are observed. When the particle starts oscillating a second strong peak at 25 Hz appears. The comparison between the particle position obtained from the video and the back signal, shows clear dips of the back signal when the particle changes its metastable position. 

The observed influence of particle trapping on the back signal is very useful for future nanoparticle trapping experiments. The back signal can thus replace the direct optical visualization which is impossible for nanoparticles. 

\subsection{Theoretical description of trapping efficiency}

There are three main models to deduce the trapping efficiency from trapped particle position fluctuations. They are related to different fluctuation properties: the analysis of position oscillation power spectra, the position autocorrelation, and the position statistics. For sake of clarity, the models will be described for a motion along only one dimension ($x$). Their generalization to two or three dimensions is straightforward. 

\begin{figure}
	\centering
	\includegraphics[width=6.cm]{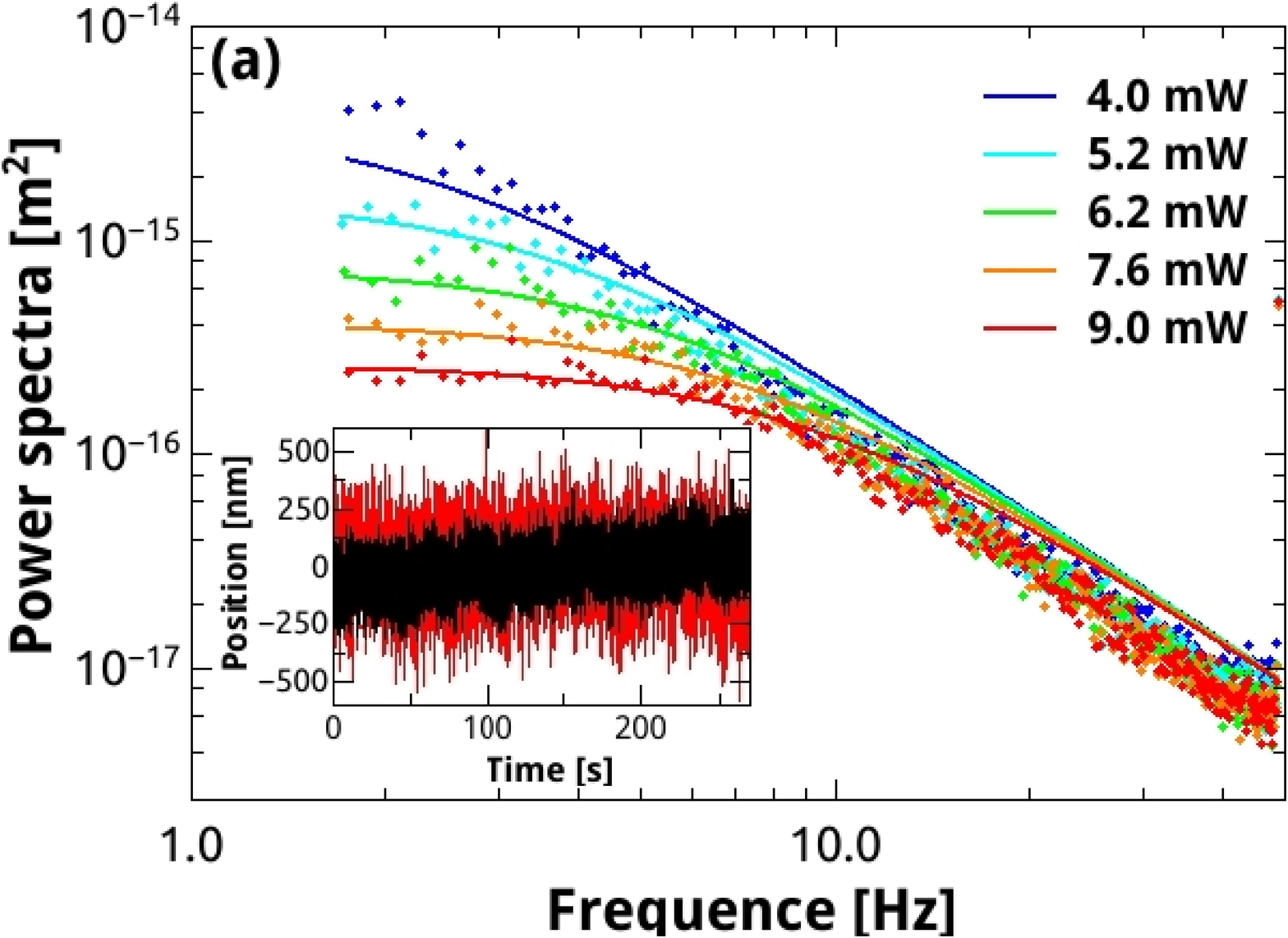}
	\includegraphics[width=6.cm]{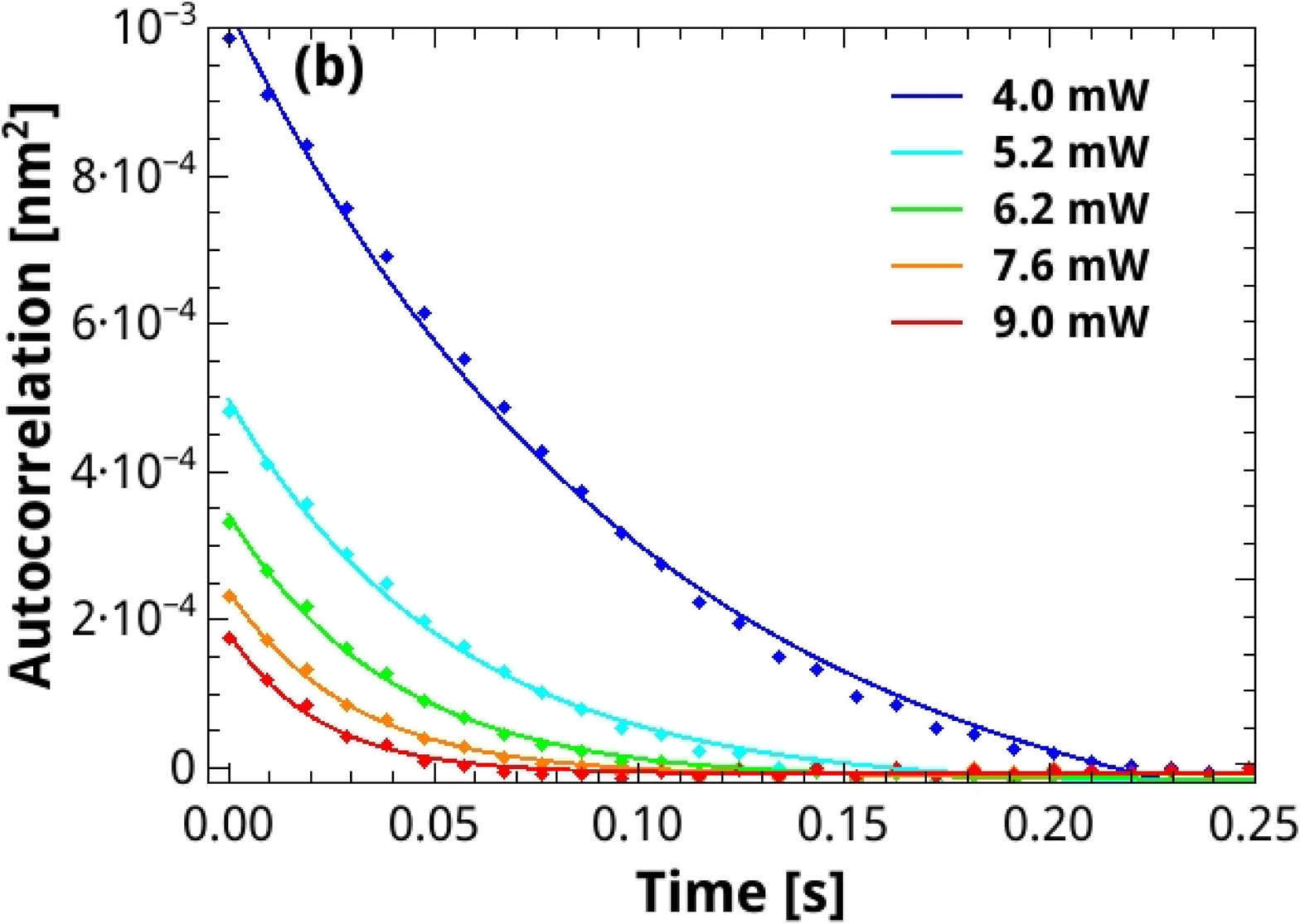}
	\caption{Power spectra (a) and autocorrelation (b) of transversal position of a trapped particle for different light intensities ($d=11.5 \mu$m). Lines are best fits to the experimental data. Insert: plot of the particle position fluctuations.\label{fig.ec}}
\end{figure}

The motion of a sphere in a harmonic trapping potential can be described by \cite{BF04}:

\begin{equation}\label{eq.eou}
   m\ddot x(t)+\gamma_0\dot x(t)+\kappa x(t)=(2k_BT\gamma_0)^{1/2}\xi(t)
\end{equation}

with $x(t)$ the trajectory of the Brownian particle, $m$ its mass, $a$ its radius, $\gamma_0=6\pi\eta a$ the friction coefficient deduced from Stokes's law, $-\kappa x(t)$ the harmonic force from the trap and $(2k_BT\gamma_0)^{1/2}\xi(t)$ the Brownian force at temperature T. The characteristic time for loss of kinetic energy through friction ($m/\gamma_0\approx6\times10^{-8}$ s) is much shorter that our experimental time resolution at 100 Hz sampling rate. Consequently we can neglect the first term in Eq. \ref{eq.eou}. Performing a Fourier transform, the position power spectrum can then be approximated by the Lorentzian:

\begin{equation}\label{eq.lor}
   P_k=\frac{2k_bT}{\gamma_0(f_c^2+f_k^2)}
\end{equation}

with $f_c=\kappa /2\pi \gamma_0$ the corner frequency and $f_k$ the oscillation frequency from the Fourier transform. The power spectra shows two distinct regimes [Fig. \ref{fig.ec}(a)]. For frequencies below the corner frequency ($f_k \ll f_c$) the power spectra is constant ($P^{\Downarrow }=8\pi^2k_BT\gamma_0/\kappa^2$) and directly dependent on the trapping stiffness $\kappa$. At high frequencies ($f_k \gg f_c$) the Brownian motion is dominant and the power spectrum is independent from $\kappa$ ($P^{\Uparrow}=2k_BT/\gamma_0 f_k^2$).

These two regimes are also visible in the autocorrelation function. In fact the residual Brownian motion $\langle x^2\rangle$ of the particle is given by the equipartition of energy \cite{BF04,GLK+08}:

\begin{equation}
  \frac{1}{2} k_BT=\frac{1}{2}\kappa\langle x^2\rangle.
\end{equation}

Since the oscillator is significantly over damped, the autocorrelation of the particle position is described by a single exponential decay of time constant $\tau_0=1/2\pi f_c$ [Fig. \ref{fig.ec}(b)].

\begin{figure}
    \centering
		\includegraphics[width=6.cm,height=4.5cm]{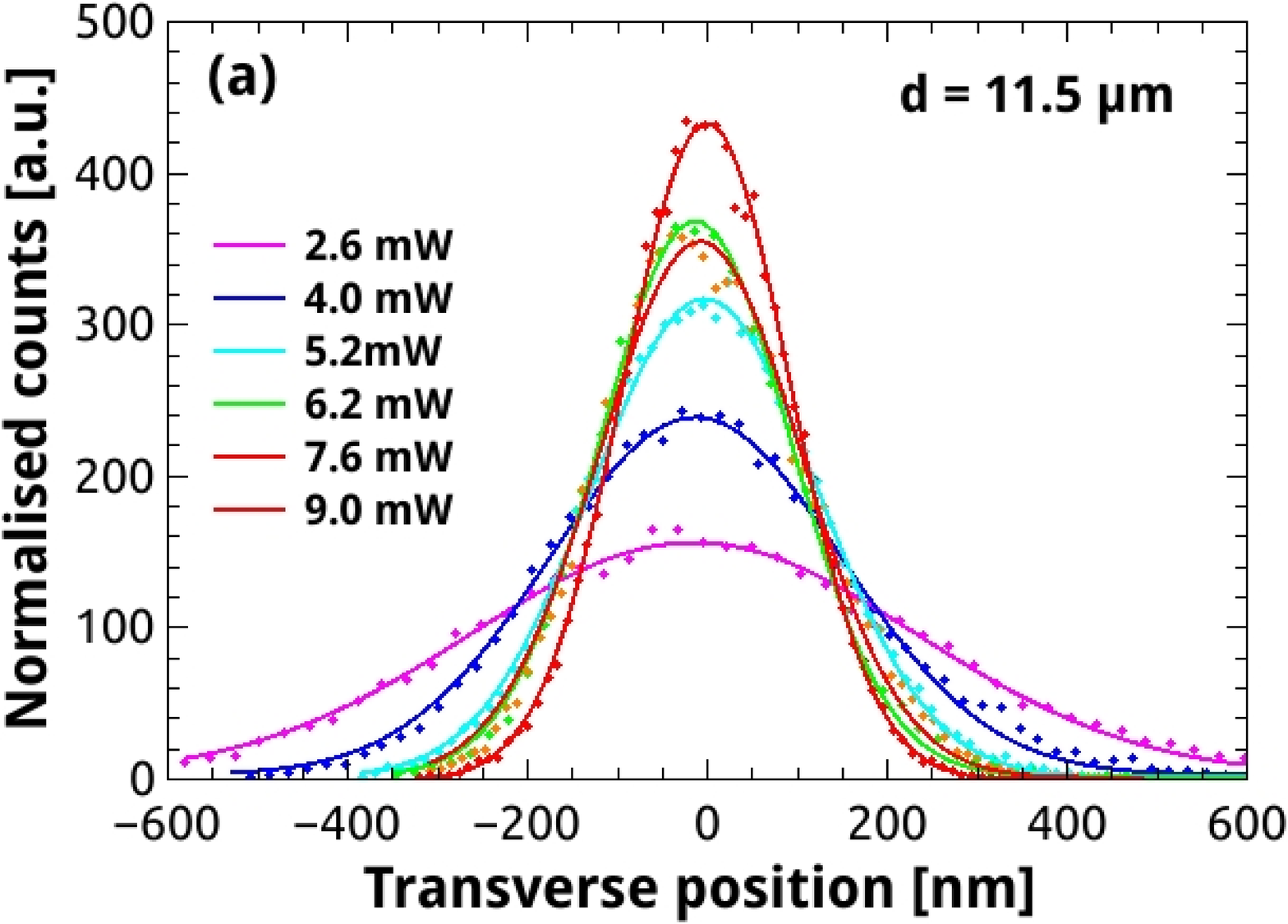}
      \includegraphics[width=6.cm,height=4.5cm]{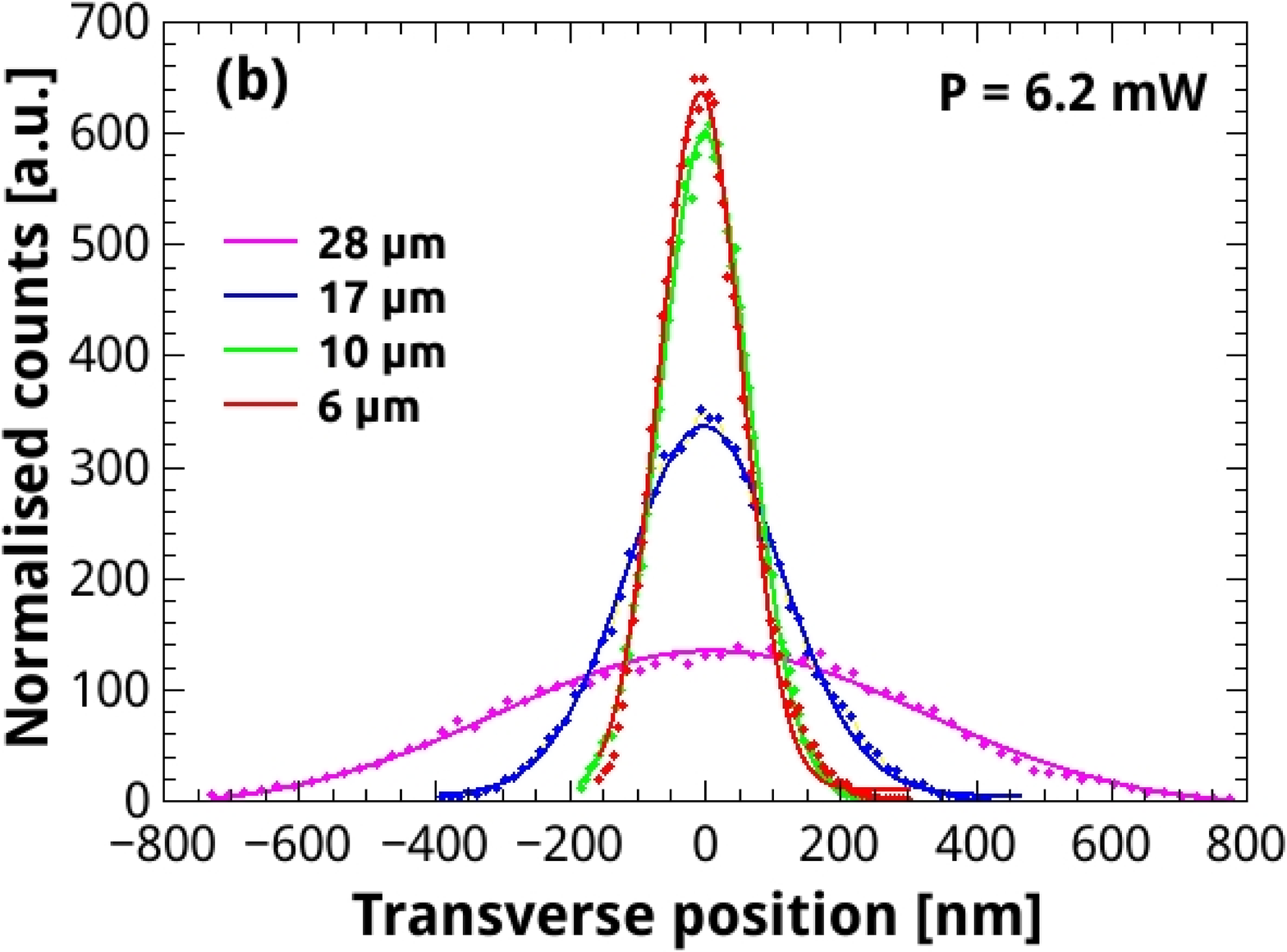}\\
      \includegraphics[width=6.cm,height=4.5cm]{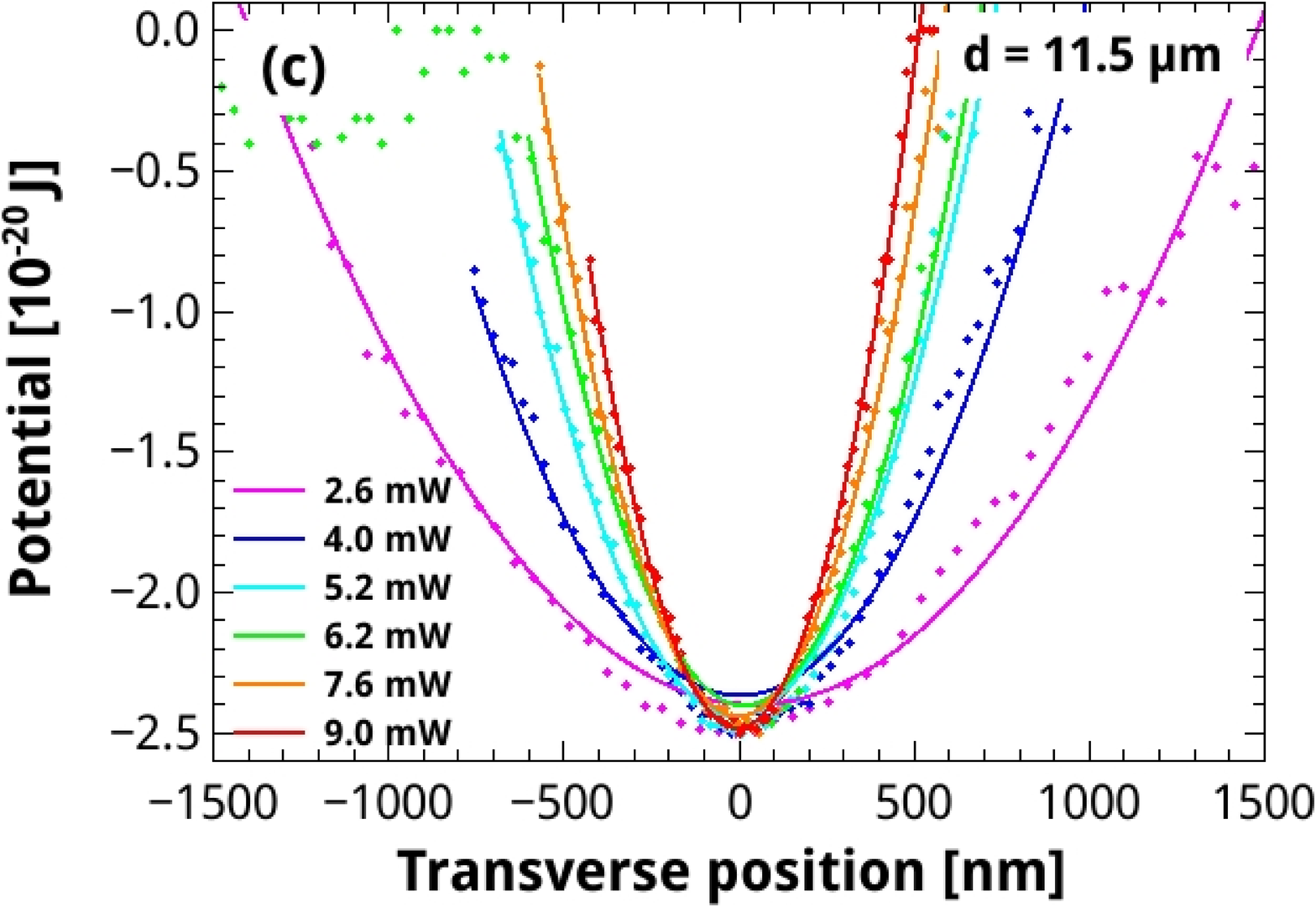}
      \includegraphics[width=6.cm,height=4.5cm]{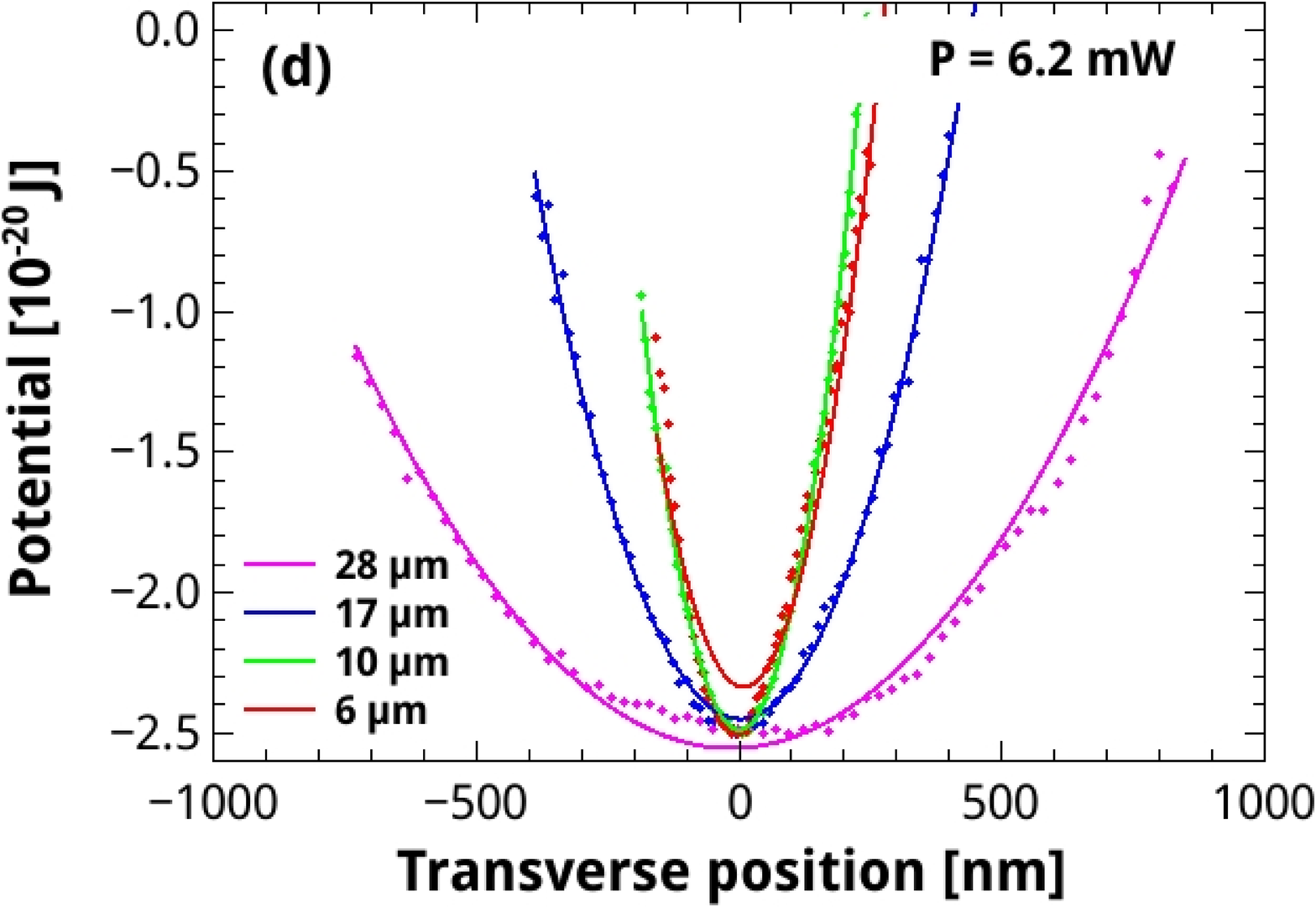}\\

    \caption{Transverse position distribution of the trapped  particle for different light intensities (a) and fiber tip-to-tip distances (b) and the corresponding trap potentials (c),(d). \label{fig.prob}}
\end{figure}

Finally, the probability density $P$ of finding the particle in the potential well $U$ at a certain position $x$ can be described using Boltzmann statistics \cite{TSS12}: 

\begin{equation}
P(x)=\frac{1}{Z}e^{\frac{-U(x)}{k_B T}}
\end{equation}

with $Z$ the partition function. For harmonic trapping potentials the trap stiffness can be directly obtained by fitting the probability density to the Gaussian function $P(r)=exp(-\kappa x^2/2k_B T)$ [Fig. \ref{fig.prob}]. This last model is the most versatile, as it is not restricted to harmonic potentials. Moreover it gives direct access to the trapping potential. This point is of great interest for traps with multiple (meta-)stable trapping positions.

The first two models are based on a frequency analysis of the particle position fluctuations. Valuable results require position detection at frequencies above the corner frequency $f_c$ of the trapped particle. In the present case $f_c$ is below 10 Hz. The CMOS camera readout frequency of $\approx100$ fps is thus sufficient. For higher trap stiffnesses this condition can, however, be a serious limitation.   

\subsection{Experimental trapping efficiencies}

The experimental particle position fluctuations are obtained using our tracking program. Videos of typical length of 5 minutes and over $3\times 10^5$ frames ensure good statistics [insert Fig. \ref{fig.ec}(a)]. The position for the transverse and longitudinal positions are determined separately, thus allowing to calculate the corresponding forces independently.  

Two series of trapping experiments are conducted in one go using a pair of identical fiber tips. The first series is recorded at a fixed fiber distance of 11.5 $\mu$m and light powers between 2.6 and 9.0 mW. The seconds series is recorded at fixed power of 6.2 mW and fiber tip-to-tip distances of 6 to 28 $\mu$m. Their transverse probability density functions are displayed on Figs. \ref{fig.prob}(a)--\ref{fig.prob}(b). The experimental curves fits very well to the Gaussian function.

\begin{figure}
	\centering
	\includegraphics[width=6.cm]{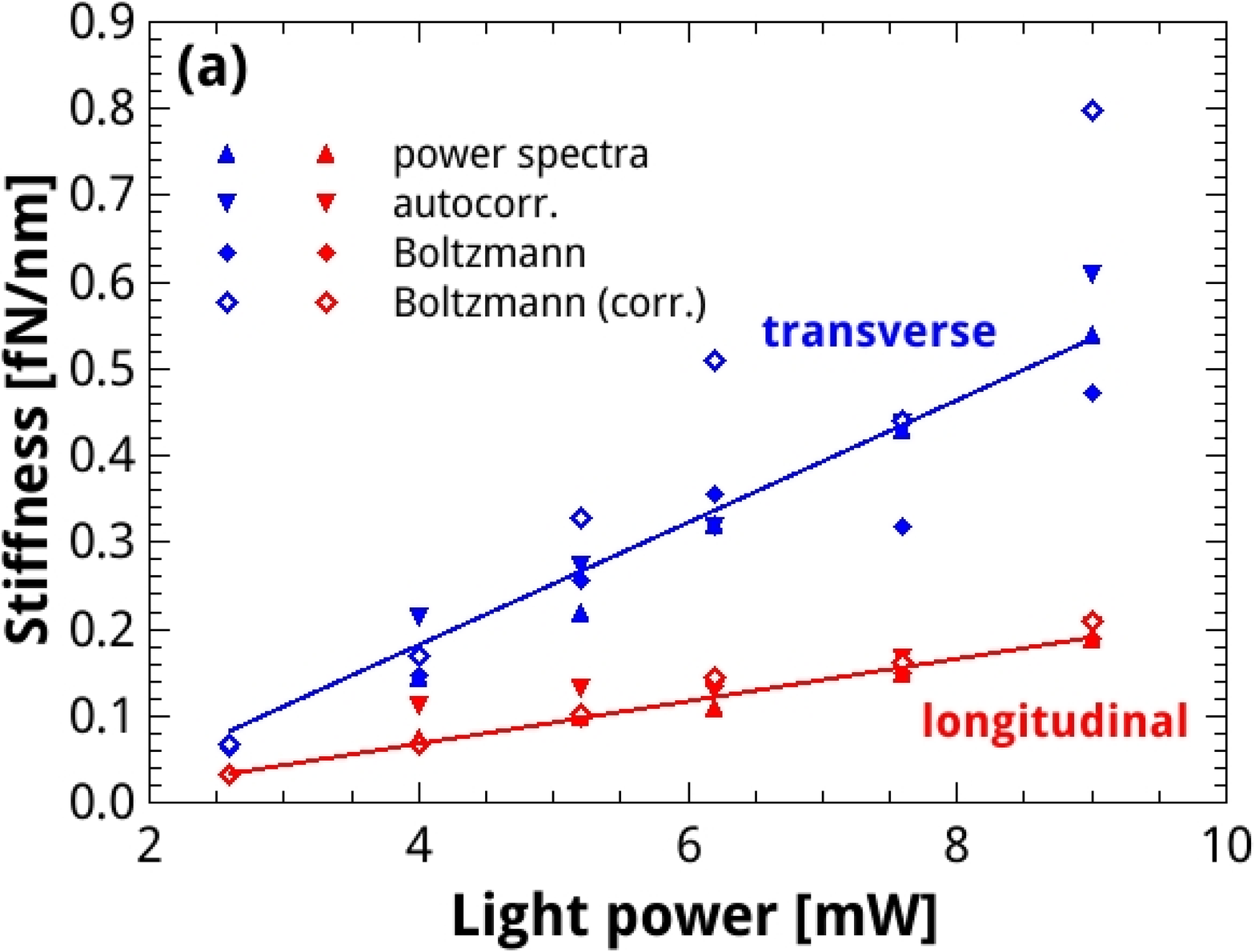}
	\includegraphics[width=6.cm]{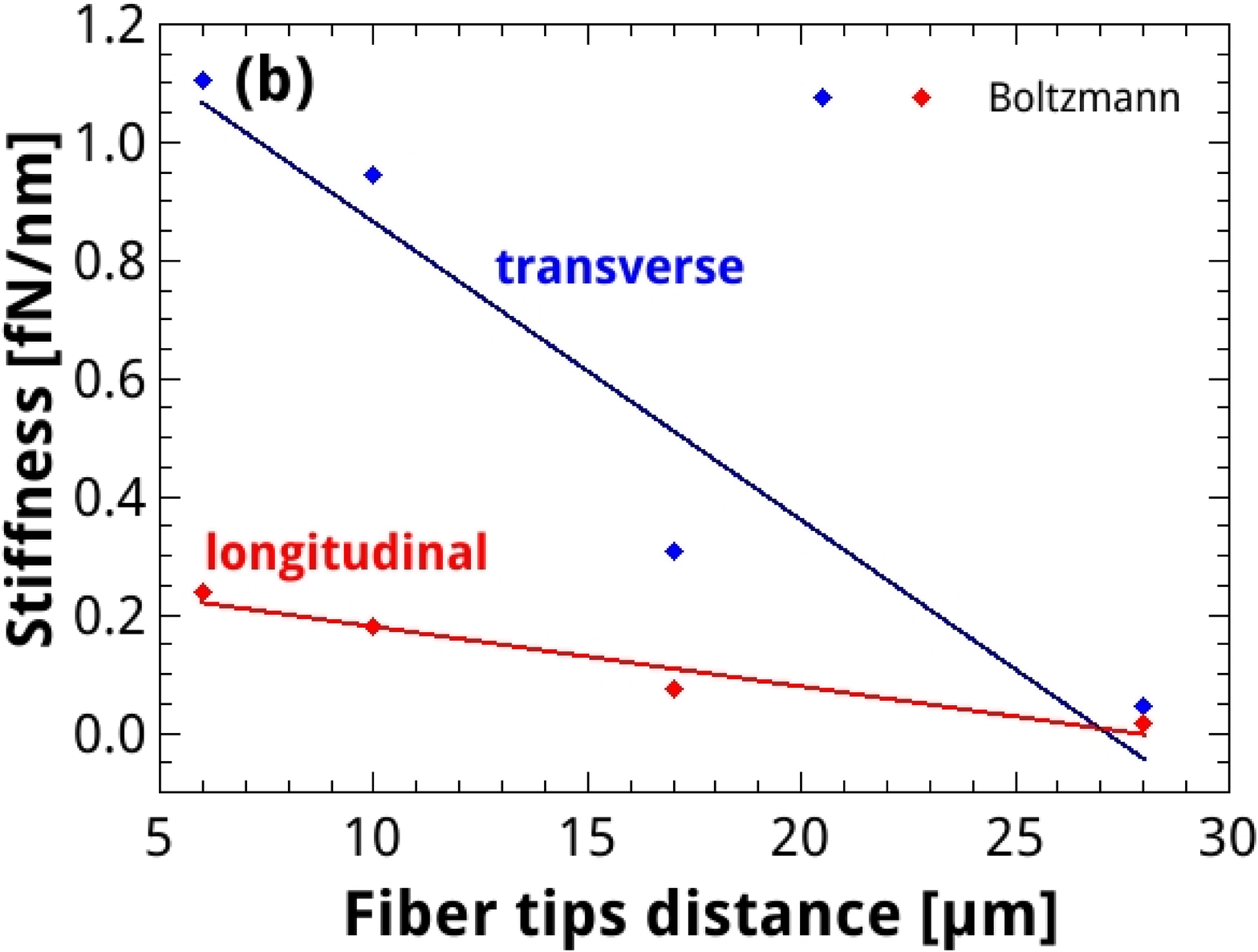}
	\caption{Trap stiffness along transversal and longitudinal directions as a function of laser power at the end of each fibers (a) and as a function of fiber tip-to-tip distance (b).}
	\label{fig.stiff}
\end{figure}

Depending on light intensity and tip-to-tip distance, the calculated trap stiffnesses $\kappa_t$ are of the order 0.1 to 1 fN/nm and 0.05 to 0.2 fN/nm for the transverse and longitudinal direction, respectively [Fig. \ref{fig.stiff}]. The trapping forces increase linearly with increasing light intensity and decreasing tip-to-tip distance. The important difference between the longitudinal and transversal trapping forces is due to the non-conformity of the optical trap.  The scattering forces from the two counter-propagating beams push the particle in two opposite directions whereas the gradient forces pull it into the same direction. Consequently the sphere is strongly stabilized in the transverse direction. 

A control video of a particle fixed on a substrate is recorded to evaluate the noise of the experimental set. The obtained probability distribution is of Gaussian shape with a width of the order of 50 nm. This value is used for the correction of the measured trap stiffness [Fig. \ref{fig.stiff}]. The correction becomes significant for higher power transverse trapping, with its narrow position distribution. 

The potential wells calculated from the position distributions are shown on Fig. \ref{fig.prob}. With the exception of the low intensity case, the potentials are of clear parabolic shape. This point justifies the subsequent application of the power spectra and autocorrelation models for the determination of the trapping forces. The results of these models will be presented only for the series with a fixed tip-to-tip distance. 

The autocorrelation functions of the transverse position fluctuations are shown on Fig. \ref{fig.ec}(b). The corner frequencies $f_c$ are obtained by fitting to single exponential decay curves. The autocorrelation function decreases faster for stronger trapping with, for example, $f_c= 7.07$ and 1.53 Hz for light intensities of 9.0 and 2.6 mW respectively. The corresponding trapping stiffnesses are calculated using $\gamma_0=9.44 \times 10^{-9}$ Ns/m for 1 $\mu$m particles in water at room temperature [Fig. \ref{fig.stiff}]. 

Finally, the transverse spectral densities are plotted on Fig. \ref{fig.ec}(a). The agreement with the best numerical fit to Eq. \ref{eq.eou} is very good in the low frequency regime. A slight shift is, however, observed for frequencies above the corner frequency. As stated before, our model comprises any free fitting parameter in this frequency regime. We suggest that the observed discrepancy is caused by the limited performances of our CMOS camera at these frequencies. The $\kappa$ values obtained by the low frequency fits are includes in Fig. \ref{fig.stiff}(a). 

The agreement of the values deduced by the three presented models is very good for the weaker longitudinal trapping direction. The agreement remains satisfactory in the transverse direction. In this case the correction made for Boltzmann model becomes, however, significant.

\section{Conclusions}

Optical fiber tips with nanometer size apex are used for trapping experiments in single fiber and dual fiber optical tweezers. In the single fiber tip geometry, 1 $\mu$m size dielectric particles in water are trapped for some seconds before being ejected by the scattering forces of about 0.4 -- 0.7 fN. Stable trapping of the same particles is observed in the dual fiber tip tweezers for light intensities down to 2.6 mW. Trap stiffnesses of up to 1 fN/nm are deduced from particle position fluctuation traces using three different theoretical approaches. The results of the three models are in very good agreement. The Boltzmann statistics model is, however, found to be the most appropriate as it allows to deal with non-harmonic traps and to include corrections for experimental set-up noise. The trap stiffness is about 2.5 times higher in the transverse direction than in the longitudinal direction. It is linearly increasing with light intensities and decreasing with tip-to-tip distance.   

The presented results are promising for future nanoparticle trapping experiments. In this respect, preliminary trapping of single fluorescent YAG particles \cite{RDD+11, MMB+13} of deep sub-micron size with our dual nano-tip tweezers confirms the potential of this trapping scheme.

\section*{Acknowledgments}
Funding for this project was provided by the French National Research Agency in the framework of the FiPlaNT project (ANR-12-BS10-002). Authors thank Jean-Fran\c{c}ois Motte for the elaboration of the applied fiber tips. Helpful discussions with G. Colas des Francs, G. Dantelle, and T. Gacoin are gratefully acknowledged.
 

\begin{thebibliography}{10}
\newcommand{\enquote}[1]{``#1''}

\bibitem{ADB+86}
A.~Ashkin, J.~M. Dziedzic, J.~E. Bjorkholm, and S.~Chu, \enquote{Observation of
  a single-beam gradient force optical trap for dielectric particles,} Opt.
  Lett. \textbf{11}, 288--290 (1986).

\bibitem{HLS+12}
O.~G. Helleso, P.~Lovhaugen, A.~Z. Subramanian, J.~S. Wilkinson, and B.~S.
  Ahluwalia, \enquote{Surface transport and stable trapping of particles and
  cells by an optical waveguide loop,} Lab Chip \textbf{12}, 3436--3440 (2012).

\bibitem{RDC+12}
C.~Renaut, J.~Dellinger, B.~Cluzel, T.~Honegger, D.~Peyrade, E.~Picard, F.~d.
  Fornel, and E.~Hadji, \enquote{Assembly of microparticles by optical trapping
  with a photonic crystal nanocavity,} Appl. Phys. Lett. \textbf{100}, 101103
  (2012).

\bibitem{WSS+10}
K.~Wang, E.~Schonbrun, P.~Steinvurzel, and K.~B. Crozier, \enquote{Scannable
  plasmonic trapping using a gold stripe,} Nano Lett. \textbf{10},
  3506--3511 (2010).

\bibitem{ZLS+10}
W.~Zhang, L.~Lina~Huang, C.~Santschi, and O.~J.~F. Martin, \enquote{Trapping
  and sensing 10 nm metal nanoparticles using plasmonic dipole
  antennas,} Nano Lett. \textbf{10}, 1006--1011 (2010).

\bibitem{TS11}
Y.~Tanaka and K.~Sasaki, \enquote{Optical trapping through the localized
  surface-plasmon resonance of engineered gold nanoblock pairs,} Opt. Express
  \textbf{19}, 17462--17468 (2011).

\bibitem{PG11}
Y.~Pang and R.~Gordon, \enquote{Optical trapping of a single protein,}
  Nano Lett. \textbf{12}, 402--406 (2011).

\bibitem{BLM12}
J.~B. Black, D.~Luo, and S.~K. Mohanty, \enquote{Fiber-optic rotation of
  micro-scale structures enabled microfluidic actuation and self-scanning
  two-photon excitation,} Appl. Phys. Lett. \textbf{101}, 221105 (2012).

\bibitem{VOO09}
S.~Valkai, L.~Oroszi, and P.~Ormos, \enquote{Optical tweezers with tips grown
  at the end of fibers by photopolymerization,} Appl. Optics \textbf{48},
  2880--2883 (2009).

\bibitem{LS95}
E.~R. Lyons and G.~J. Sonek, \enquote{Confinement and bistability in a tapered
  hemispherically lensed optical fiber trap,} Appl. Phys. Lett. \textbf{66},
  1584--1586 (1995).

\bibitem{NTO+06}
T.~Numata, A.~Takayanagi, Y.~Otani, and N.~Umeda, \enquote{Manipulation of
  metal nanoparticles using fiber-optic laser tweezers with a
  microspherical focusing lens,} Japanese J. Appl. Phys. \textbf{45},
  359--363 (2006).

\bibitem{BKA+13}
A.~L. Barron, A.~K. Kar, T.~J. Aspray, A.~J. Waddie, M.~R. Aghizadeh, and H.~T.
  Bookey, \enquote{Two dimensional interferometric optical trapping of multiple
  particles and {E}scherichia coli bacterial cells using a lensed multicore
  fiber,} Opt. Express \textbf{21}, 13199--13207 (2013).

\bibitem{LGY+06}
Z.~Liu, C.~Guo, J.~Yang, and L.~Yuan, \enquote{Tapered fiber optical tweezers
  for microscopic particle trapping: fabrication and application,} Opt. Express
  \textbf{14}, 12510--12516 (2006).

\bibitem{LWL+13}
Z.~Liu, L.~Wang, P.~Liang, Y.~Zhang, J.~Yang, and L.~Yuan, \enquote{Mode
  division multiplexing technology for single-fiber optical trapping
  axial-position adjustment,} Opt. Lett. \textbf{38}, 2617--2620 (2013).

\bibitem{MPK12}
S.~K. Mondal, S.~S. Pal, and P.~Kapur, \enquote{Optical fiber nano-tip and 3{D}
  bottle beam as non-plasmonic optical tweezers,} Opt. Express \textbf{20},
  16180--16185 (2012).

\bibitem{MHN+10}
M.~Michihata, T.~Hayashi, D.~Nakai, and Y.~Takaya, \enquote{Microdisplacement
  sensor using an optically trapped microprobe based on the interference
  scale,} Rev. Sci. Instrum. \textbf{81}, 015107 (2010).

\bibitem{BF04}
K.~Berg-S\o{}rensen and H.~Flyvberg, \enquote{Power spectrum analysis for
  optical tweezers,} Rev. Sci. Instrum. \textbf{75}, 594--612 (2004).

\bibitem{GLK+08}
G.~M. Gibson, J.~Leach, S.~Keen, A.~J. Wright, and M.~J. Padgett,
  \enquote{Measuring the accuracy of particle position and force in optical
  tweezers using high-speed video microscopy,} Opt. Express \textbf{16},
  14561--14570 (2008).

\bibitem{DSV+11}
J.-B. Decombe, W.~Schwartz, C.~Villard, H.~Guillou, J.~Chevrier, S.~Huant, and
  J.~Fick, \enquote{Living cell imaging by far-field fibered interference
  scanning optical microscopy,} Opt. Express \textbf{19}, 2702--2710 (2011).

\bibitem{CSM+06}
N.~Chevalier, Y.~Sonnefraud, J.~F. Motte, S.~Huant, and K.~Karrai,
  \enquote{Aperture-size-controlled optical fiber tips for high-resolution
  optical microscopy,} Rev. Sci. Instrum. \textbf{77}, 063704 (2006).

\bibitem{DBM+13}
J.~B. Decombe, J.~F. Bryche, J.~F. Motte, J.~Chevrier, S.~Huant, and J.~Fick,
  \enquote{Transmission and reflection characteristics of metal-coated optical
  fiber tip pairs,} Appl. Optics \textbf{52}, 6620--6625 (2013).

\bibitem{TSS12}
Y.~Tanaka, A.~Sanada, and K.~Sasaki, \enquote{Nanoscale interference patterns
  of gap-mode multipolar plasmonic fields,} Sci. Rep. \textbf{2}, 764 (2012).

\bibitem{RDD+11}
A.~Reveaux, G.~Dantelle, D.~Decanini, A.-M. Haghiri-Gosnet, T.~Gacoin, and
  J.-P. Boilot, \enquote{Synthesis of {YAG}:{C}e/{T}i{O}2 nanocomposite films,}
  Opt. Mater. \textbf{33}, 1124--1127 (2011).

\bibitem{MMB+13}
B.~Masenelli, O.~Mollet, O.~Boisron, B.~Canut, G.~Ledoux, J.-M. Bluet,
  P.~M\'elinon, C.~Dujardin, and S.~Huant, \enquote{Y{AG}:{C}e nanoparticle
  lightsources,} Nanotechnology \textbf{24}, 165703 (2013).

\end{thebibliography}

\end{document}